\pdfoutput=1
\documentclass[onecolumn]{techreport}
\usepackage{listings}
\usepackage{longtable}
\usepackage{array}
\usepackage{tabularx}
\usepackage{makecell}
\usepackage{tikz}
\usepackage{footmisc}
\DeclareMathOperator*{\argmax}{\arg\!\max}
\DeclareMathOperator*{\argmin}{\arg\!\min}
\newcommand{\R}{\mathbb{R}}

\title{A thorough review on Deep Learning approaches to 2D-to-3D Registration in Medical Imaging}

\author{
\begin{center}
  Andrea Borghesi \quad Julia van Beurden \quad Maarten van Alphen \quad Michiel van den Brekel \\[2pt]
  Luc Karssemakers \quad Jan-Jakob Sonke \quad Baris Karakullukcu \\[2pt]
  Jonas Teuwen$^{\dagger}$ \quad George Yiasemis$^{\dagger}$ \\[12pt]
  $^{\dagger}$Corresponding authors
\end{center}
}

\begin{document}
\abstract{%
2D-to-3D image registration is a key component for several medical imaging purposes such as intraoperative guidance, enabling the alignment between preoperative 3D volumes and real-time 2D images. While conventional optimisation-based methods often struggle with non-convex objectives, sensitivity to initialisation, and high computational demands, deep learning (DL) has emerged as a promising alternative, offering improved alignment robustness and the potential for real-time performance. 
This review provides a comprehensive analysis of DL frameworks developed to address the challenges inherent in 2D-to-3D registration. 
We first outline how core concepts of DL are adapted to the registration problem. We then examine recent methodological advancements from multiple perspectives, including pipeline design, architectural innovations, and domain adaptation strategies. Particular attention is given to quantitative metrics, highlighting how performance is defined and measured across studies. 
Across the 41 included studies, we find that direct regression is the most common pipeline (23 studies, against 18 iterative ones), that hybrid CNN-attention architectures are gaining ground, and that research remains heavily concentrated on X-ray-to-CT registration. 
We conclude by identifying the absence of per-case uncertainty estimation and the strong neglect of temporal or sequential modelling as the two methodological gaps most consequential for clinical readiness, and discuss state-space models as a way to add temporal memory within intraoperative latency budgets. Further opportunities lie in more principled domain adaptation and in the integration of foundation models for representation learning.
}
\maketitle

\section{Introduction}

Image-guided interventions increasingly combine high-resolution preoperative volumes with real-time intraoperative imaging. Computed tomography (CT) or magnetic resonance imaging (MRI) is acquired before the procedure for diagnosis, planning and target delineation, but the same volumetric modalities are rarely available during the procedure: acquisition is slow, the scanner is incompatible with most surgical equipment, CT adds radiation dose, and either disrupts the clinical workflow. Intraoperative imaging is therefore predominantly two-dimensional. X-ray fluoroscopy and single cone-beam projections guide spinal, pelvic and endovascular procedures and verify patient position in radiotherapy \cite{uneriDeformable3D2DRegistration2021, chinaAutomatedLabelingUsing2025, downsImprovingAutomaticCerebral2025, yeNonorthogonalKVImaging2025, moriShorteningImageRegistration2023}; ultrasound (US) guides biopsies, thermal ablations and abdominal surgery \cite{wood2005navigation, krucker2007electromagnetic, venkatesan2011real, hiepRealtimeIntraoperativeUltrasound2025, guoUltrasoundFrametoVolumeRegistration2023a}; endoscopic and laparoscopic video is the only view available in minimally invasive surgery \cite{kobayashiShapeRegistrationLaparoscopic2025, oya2D3DDeformable2024}. These 2D modalities are fast, inexpensive, portable and, where ionising radiation is involved, deliver a lower dose per image than a volumetric scan. Their limitation is that they show a projection or a single plane of a three-dimensional anatomy, so the clinician must mentally map each 2D image onto the 3D plan.

This mental mapping is error-prone, and the errors are costly. In a survey of spine surgeons, half of the respondents reported having operated on the wrong vertebral level at least once in their career, with an estimated prevalence of one wrong-level operation per 3\,110 procedures \cite{mody2008prevalence}; a national survey of neurosurgeons identified the failure to verify the operative site radiographically as a common contributing factor \cite{jhawar2007wrong}. The usual safeguard is to verify the operative level with additional radiographs or fluoroscopic images of the same patient during the procedure, which lowers the risk of a wrong-level operation but raises the radiation exposure of the patient, the surgeon and the operating-room staff. Fluoroscopically assisted pedicle screw placement exposes the surgeon to measurable radiation at every acquisition \cite{rampersaud2000radiation}, minimally invasive spine procedures require substantially more fluoroscopy than open ones \cite{srinivasan2014radiation}, and an increased cancer incidence has been reported among orthopaedic surgeons working with inadequate radiation protection \cite{mastrangelo2005increased}. In image-guided radiotherapy, daily positioning images add a cumulative dose that must be managed explicitly \cite{murphy2007management}. Anatomy is also not static between planning and treatment: respiratory motion displaces thoracic and abdominal targets \cite{keall2006management}, causing under-dosage of the tumour and over-dosage of organs at risk (OARs), and organs deform under insufflation and surgical manipulation \cite{caoNonrigidImagevolumeRegistration2025, mhiriNeuralPatientspecific3D2D2025b}. Automatically aligning each intraoperative 2D image with the preoperative 3D volume, quickly and accurately, addresses both problems at once: it removes the need for mental reconstruction and it extracts the required 3D information from images that are already acquired. This is the task of 2D-to-3D image registration.

Most of the studies included in this review register fluoroscopic or radiographic images to CT or CBCT, to guide spinal, orthopaedic and endovascular procedures or to position patients and track tumours in radiotherapy \cite{miaoCNNRegressionApproach2016, quTransformerbased2D3D2024b, wuDVAPRegDualviewAnatomical2026, zhangEnhancingCNNRegressors2026, wangPoseawareDeepPerceptual2026, downsImprovingAutomaticCerebral2025, daiVolumetricTumorTracking2024, yeNonorthogonalKVImaging2025, hindleyOpenSourceDeep2025}. Others localise the US probe in a preoperative volume during prostate, kidney and liver interventions \cite{dupuy2D3DDeep2023a, soormallyDeepSensorlessTracking2025, guoUltrasoundFrametoVolumeRegistration2023a, chiTwostepDeepLearning2023, zhangUltrasoundSlicetoVolumeRegistration2026}, or overlay a preoperative organ model on the laparoscopic or thoracoscopic view \cite{kobayashiShapeRegistrationLaparoscopic2025, mhiriNeuralPatientspecific3D2D2025b, zhangWarmstartedReinforcementLearning2026, caoNonrigidImagevolumeRegistration2025, oya2D3DDeformable2024}. The same problem also arises offline, when a high-resolution fetal brain volume is reconstructed from motion-corrupted MRI slices \cite{xuSVoRTIterativeTransformer2022, wu3DIsotropicHighResolution2025} or a patient-specific bone model from standard radiographs \cite{khamenehAutomaticCNNbased3D2025, jiangSemanticallyConsistentDeformable2D3D2025a}.

Image registration is the estimation of the spatial transformation that aligns two images in a common coordinate system so that anatomical structures can be compared, fused or tracked. When both images have the same dimensionality (2D-to-2D or 3D-to-3D), corresponding structures share the same spatial representation and the geometric relationship between them is relatively direct. 2D-to-3D registration adds a dimensional mismatch. For example a radiograph integrates the anatomy along each projection ray, and an US, MRI or endoscopic frame samples a single plane or surface, so a 2D image can only be compared with the volume after the volume has been projected or sliced under a candidate transformation, an operation that discards volumetric context and leaves the displacement along the projection or through-plane axis poorly constrained. The two images typically also come from different physical processes (X-ray versus CT, US versus CT or MRI, optical camera versus CT), so their intensities are not directly comparable, and the anatomy may have deformed between the two acquisitions. Finally, intraoperative use imposes strict time budgets, often a fraction of a second per frame.

Early efforts in 2D-to-3D registration relied on iterative optimisation of hand-crafted objective functions, as reviewed by \citet{markelj2012review}, and are generally classified as feature-based or intensity-based. Feature-based methods reduce the images to geometric primitives such as points, lines or contours and minimise the distance between corresponding structures; representative examples are the Iterative Closest Point algorithm \cite{besl1992method} and its variants matching 3D surface meshes to 2D contours \cite{benseghir2013iterative, yavariabdi2015mapping, fitzgibbon2003robust}, landmark-based registration with fiducial or anatomical points \cite{sahovaler2022automatic}, and matching of bone contours or silhouettes \cite{lavallee1995recovering, gueziec1998anatomy}. They are computationally efficient but depend on the quality of segmentation and on the visibility of landmarks. Intensity-based methods instead use the pixel and voxel values directly: digitally reconstructed radiographs (DRRs) are generated from the volume by ray casting and compared with the real X-ray image through similarity metrics such as mutual information (MI) \cite{zollei2001mutual}, normalised cross-correlation (NCC) or gradient correlation \cite{penney1998comparison}. The resulting objective is highly non-convex. Local optimisers such as Powell's method are efficient but require an accurate initial pose to converge \cite{miaoCNNRegressionApproach2016}, whereas global strategies such as evolutionary algorithms (CMA-ES) tolerate an unknown initial pose at a computational cost that is prohibitive for real-time use \cite{otake2016robust}.

Deep learning (DL) has emerged as a way out of this trade-off. A network can be trained to predict the transformation directly from the image pair, or to refine it in a small number of learned steps, shifting the computational burden from the intraoperative inference phase to offline training. Learned methods enlarge the capture range, so that accurate alignment is achieved from poor initialisations, and reach inference times compatible with real-time guidance. They have also enabled new formulations, such as self-supervised training through differentiable projection operators and test-time optimisation initialised by a network prediction, that blur the line between learning and classical optimisation. Despite this rapid growth, existing reviews either situate 2D-to-3D methods within broader surveys of image registration \cite{chenSurveyDeepLearning2025}, thereby underemphasising the specific projective-to-volumetric challenges, or predate most recent developments \cite{unberathImpactMachineLearning2021}.

This review therefore addresses the question: \textit{what deep learning frameworks have been developed to overcome the challenges of 2D-to-3D medical image registration in clinical environments?} We answer it through a systematic review, conducted according to the PRISMA 2020 guidelines, of the peer-reviewed literature on the topic. Its contributions are (i) a unified formulation of classical, supervised and self-supervised 2D-to-3D registration that fixes the notation used throughout the paper; (ii) a structured analysis of the included studies along three axes, the learning paradigm, the pipeline design (direct regression versus iterative refinement) and the architectural components, including differentiable geometric operators, generative models and implicit neural representations; (iii) a critical examination of how performance is measured and on which data, including a per-study summary of the metrics used and of whether training and evaluation relied on synthetic or clinical images; and (iv) the identification of the gaps that most limit clinical readiness, in particular the absence of per-case uncertainty estimation and of temporal modelling, together with concrete directions on domain adaptation and foundation models. Section~\ref{sec:methods} describes the search methodology, Section~\ref{sec:problem-formulation} formalises the registration problem, Sections~\ref{sec:supvsunsup} to~\ref{sec:architectures} examine the learning paradigms, pipeline designs and architectures of the included studies, Section~\ref{sec:eval} reviews the metrics used to evaluate them, Section~\ref{sec:domain-adaptation} covers domain adaptation, Section~\ref{sec:discussion} discusses the open challenges and the limitations of this review, and Section~\ref{sec:conclusion} concludes. A \hyperref[sec:acronyms]{List of Acronyms} is given after the Conclusion.


\section{Search Methodology}
\label{sec:methods}
A structured literature search was conducted in the PubMed database to identify relevant articles published between 1 January 2015 and 30 June 2026, a period corresponding to the emergence of deep learning methods in 2D-to-3D registration. Earlier learning-based approaches to 2D/3D registration are covered by the systematic review of \citet{unberathImpactMachineLearning2021}. The search strategy was designed to capture three core concepts:
\begin{enumerate}
    \item Cross-dimensional 2D/3D images, 
    \item The image registration task, 
    \item The application of artificial intelligence-based methods. 
\end{enumerate}    
Relevant Medical Subject Headings (MeSH) terms were identified and incorporated into the search query.

The search query is reported in Appendix \ref{app:query}.

\subsection{Inclusion criteria}
Following duplicate removal, two reviewers independently screened titles and abstracts.
Eighty-four studies were deemed eligible for full-text reading and were evaluated against predefined inclusion and exclusion criteria. Disagreements were resolved through discussion; a third reviewer was consulted when consensus could not be reached.  

Studies were included if they satisfied the following criteria:
\begin{itemize}
    \item \textbf{Study design:} Original research articles proposing or evaluating methods for 2D-to-3D medical image registration.
    \item \textbf{Technology:} Primary use of deep learning for the registration task itself, rather than exclusively for preprocessing (e.g., segmentation or denoising).
    \item \textbf{Imaging modality:} Clinical 2D-to-3D modality pairs (e.g., X-ray/CT, ultrasound/MRI, optical/CT).
    \item \textbf{Language:} Articles published in English.
    \item \textbf{Format:} Full research papers; review articles, short conference abstracts, and editorials were excluded.
\end{itemize}

\begin{figure}[htbp]
    \centering
    \includegraphics[width=0.7\textwidth]{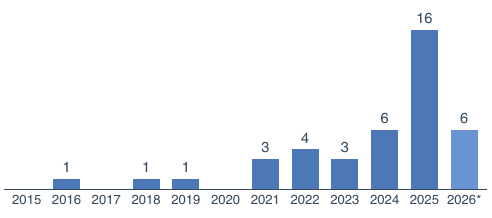}
    \caption{Number of included studies per publication year ($n=41$). Sixty-eight per cent were published from 2024 onwards. $^{*}$2026 covers January to June only, the end of the search window.}
    \label{fig:year}
\end{figure}

\begin{figure}[!htb]
    \centering
    \includegraphics[width=0.75\textwidth]{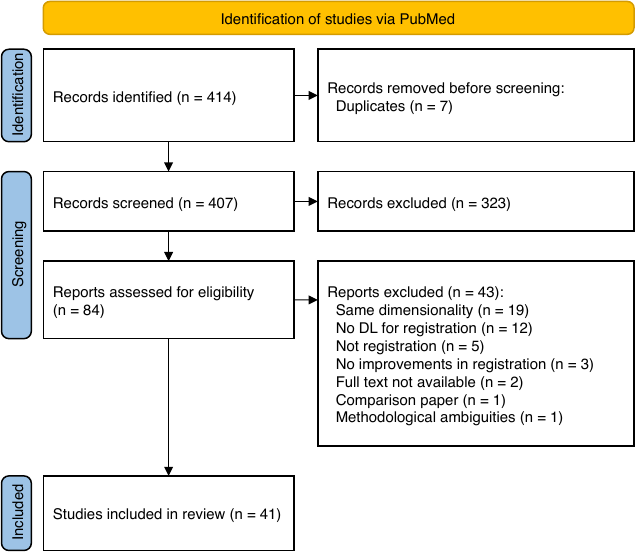}
    \caption{PRISMA 2020 flow diagram of the study selection for articles published between January 2015 and June 2026, from 414 records retrieved to 41 included studies. Exclusion reasons at each stage are given in the right-hand boxes.}
    \label{fig:prisma}
\end{figure}

The selection process is summarised in Fig.~\ref{fig:prisma}: of the 414 records retrieved, 407 remained after removing seven duplicates; 323 were excluded at title and abstract screening, and 43 of the 84 full texts assessed were excluded, most often because the registration was between images of the same dimensionality (19) or because deep learning was used only for preprocessing rather than for the registration itself (12). A total of 41 articles met all criteria and were included in the final analysis. Their temporal distribution (Fig.~\ref{fig:year}) shows that most were published from 2024 onwards (28 of 41), six of them in the first half of 2026. Each included article was assessed using a modified version of the MI-CLAIM checklist adapted for deep learning-based registration methods; detailed assessment results are provided in Appendix \ref{app:quality-assessment}.


\subsection{Data extraction}
Data extraction was performed systematically for the 41 included studies. Extracted variables were grouped into the following categories:
\begin{itemize}
    \item \textbf{General study characteristics:} Title, year of publication, first author, and DOI.
    \item \textbf{Clinical context:} Primary registration objective and target anatomy.
    \item \textbf{Imaging modalities:} Specific 2D and 3D modality pairing.
    \item \textbf{Technical architecture:} Deep learning framework, including network type, learnable parameters, and output representation.
    \item \textbf{Methodological details:} Loss functions, learning paradigm (supervised or self-supervised), and domain adaptation techniques.
    \item \textbf{Performance and validation:} Quantitative outcomes such as Target Registration Error (TRE), Mean Absolute Error (MAE), capture range, and inference time.
\end{itemize}

The resulting characteristics of all 41 included studies are summarised in Table \ref{tab:data-extraction-summary}.

\begin{table*}[!htbp]
\centering
\footnotesize
\setlength{\tabcolsep}{4pt}
\begin{tabularx}{\textwidth}{>{\raggedright\arraybackslash}l l l l l c c >{\raggedright\arraybackslash}X}
\toprule
\textbf{Article} & 
\textbf{Architecture} & 
\textbf{\makecell{Reg. \\ Task}} & 
\textbf{\makecell{2D \\ Modality}} & 
\textbf{\makecell{3D \\ Modality}} & 
\textbf{\makecell{Train. \\ data}} & 
\textbf{\makecell{Eval. \\ data}} & 
\textbf{Anatomy} \\
\midrule
    \citet{caoNonrigidImagevolumeRegistration2025} & LSTM & Deformable & Optical & CT & S & S+R & Liver \\
    \citet{chiTwostepDeepLearning2023} & CNN & Rigid & US & CT & S+R & R & Kidney \\
    \citet{daiVolumetricTumorTracking2024} & CNN/SWIN & Deformable & X-Ray & CT & S & S+P & Lung \\
    \citet{downsImprovingAutomaticCerebral2025} & CNN & Rigid & DSA & CTA & S & R & Brain \\
    \citet{dupuy2D3DDeep2023a} & CNN & Rigid & US & 3D US & S & S & Prostate \\
    \citet{facenteRealtimeAlignment3D2026} & CNN & Rigid & X-Ray & CT & S & R+P & Pelvis, Cerebral \\
    \citet{gaoFullyDifferentiableFramework2024} & CNN/Transformer & Rigid & X-Ray & CT & S & S+P & Spine \\
    \citet{gopalakrishnanIntraoperative2D3D2024} & CNN & Rigid & X-Ray & CT & S & R+P & Pelvis \\
    \citet{gopalakrishnanRapidPatientspecificNeural2025} & CNN & Rigid & X-Ray & CT & S & R+P & Pelvis, Cerebral, Cranial \\
    \citet{guoUltrasoundFrametoVolumeRegistration2023a} & CNN & Rigid & US & 3D US & R & R & Prostate \\
    \citet{hindleyOpenSourceDeep2025} & CNN & Deformable & X-Ray & CT & S & S & Lung, Heart \\
    \citet{jiangSemanticallyConsistentDeformable2D3D2025a} & CNN & Deformable & X-Ray & CBCT & S+R & S+R & Cranial \\
    \citet{khamenehAutomaticCNNbased3D2025} & CNN & Deformable & X-Ray & Atlas & S+R & R & Femur \\
    \citet{khawaledSelfattentionModelRobust2025} & CNN/Transformer & Rigid & fMRI Slice & fMRI & S & S & Cranial \\
    \citet{kobayashiShapeRegistrationLaparoscopic2025} & CNN/Diffusion & Rigid & Optical & CT & S & R & Liver \\
    \citet{lecomte-denisDomainAgnostic2D3D2025} & CNN & Deformable & X-Ray & CT & S & S & Liver \\
    \citet{liDiffusionstimulatedCTUSRegistration} & CNN/Diffusion & Rigid & US & CT & S & R+P & Abdominal \\
    \citet{mezheritskyPopulationbased3DRespiratory2022} & CNN & Deformable & US & 3D US & R & R & Liver \\
    \citet{mhiriNeuralPatientspecific3D2D2025b} & MLP & Deformable & Optical & CT & S & S+R & Liver \\
    \citet{miaoCNNRegressionApproach2016} & CNN & Rigid & X-Ray & CT & S & R+P & Knee, Distal radius, Esophagus \\
    \citet{mohsenisalehiRealTimeDeepPose2019} & CNN & Rigid & MRI Slice & MRI & S & S+R$^{*}$ & Brain (fetal) \\
    \citet{oya2D3DDeformable2024} & CNN & Deformable & Optical & CT & S & S+R & Lung \\
    \citet{qianHybridPopulationbasedPatientspecific2025} & CNN/INR & Deformable & X-Ray & CT & S & S+R & Lung \\
    \citet{quTransformerbased2D3D2024b} & GNN/SWIN & Rigid & X-Ray & CT & S & S & Spine \\
    \citet{soormallyDeepSensorlessTracking2025} & CNN & Rigid & US & 3D US & S & S & Prostate \\
    \citet{tianLiftRegLimitedAngle2022} & CNN & Deformable & X-Ray & CT & S & S & Lung \\
    \citet{wangContrastiveLearningBased2024} & CNN/SWIN & Rigid & X-Ray & CT & S & S+R$^{*}$ & Spine \\
    \citet{wangPoseawareDeepPerceptual2026} & CNN & Rigid & X-Ray & CT & S+R & R & Knee \\
    \citet{weiDeepLearningApproach2021} & CNN & Rigid & US & CT & S & S+R & Liver \\
    \citet{wu3DIsotropicHighResolution2025} & CNN & Rigid & MRI Slice & MRI & S+R & S+R$^{*}$ & Brain (fetal) \\
    \citet{wuDVAPRegDualviewAnatomical2026} & CNN/GNN & Rigid & X-Ray & CT & S+R & S+R & Spine \\
    \citet{xuSVoRTIterativeTransformer2022} & CNN/Transformer & Rigid & MRI Slice & MRI & S & S+R$^{*}$ & Brain (fetal) \\
    \citet{yeNonorthogonalKVImaging2025} & INR & Rigid & X-Ray & CT & S+P & S+P & Head and Neck \\
    \citet{yeRobustMethodReal2025} & SWIN & Rigid & X-Ray & CT & S & S & Pelvic, Cranial, Chest \\
    \citet{zhangUnsupervised2D3DDeformable2021} & CNN & Deformable & X-Ray & CT & S & S & Lung \\
    \citet{zhangDeformableRegistrationLateral2021} & CNN & Deformable & 2D Projection & CBCT & S & S+R$^{*}$ & Cranial \\
    \citet{zhangEnhancingCNNRegressors2026} & CNN & Rigid & X-Ray & CT & S & S & Spine \\
    \citet{zhangUltrasoundSlicetoVolumeRegistration2026} & CNN & Rigid & US & 3D US & S+R & S+R & Liver, Prostate \\
    \citet{zhangWarmstartedReinforcementLearning2026} & CNN/RL & Rigid & Optical & CT & S & S+R & Liver \\
    \citet{zhengPairwiseDomainAdaptation2018} & CNN & Rigid & X-Ray & CT & S+R & R & Spine \\
    \citet{zheng2022unsupervised} & CNN & Rigid & X-Ray & CT & S & S & Spine \\
\bottomrule
\end{tabularx}
\vspace{2pt}
\begin{flushleft}
\scriptsize
\textbf{Legend (Train./Eval. data):} \textbf{S}: synthetic or simulated 2D images and/or ground-truth transformations (e.g. DRRs rendered from CT, slices resampled from a volume, simulated ultrasound, randomly sampled perturbations or deformations); \textbf{R}: real clinical acquisitions from patients or volunteers; \textbf{P}: physical phantom, cadaver or animal acquisitions; for per-case optimisation methods without an offline training set, the training data is the case data itself; $^{*}$: real data used for qualitative assessment only, without quantitative ground truth.
\end{flushleft}
\caption{Overview of included studies by architecture, registration task, imaging modalities, type of data used for training and evaluation, and anatomical target.}
\label{tab:data-extraction-summary}
\end{table*}

\section{Problem Formulation}
\label{sec:problem-formulation}

\subsection{Classical Registration Formulation}
\label{sec:problem-formulation:classical}

In this section, we first formalize the general problem of image registration and then gradually refine the definition for the Deep Learning methods, distinguishing between supervised and self-supervised paradigms.
We can describe the image registration problem as the following optimization problem: 

\begin{equation}
    \label{eq:reg_general}
    \begin{gathered}
    \varphi^* = \argmax_{\varphi \in\mathcal{T}}\; \mathcal S\bigl(I_f, \mathcal W(I_m,\varphi)\bigr) - \lambda\,\mathcal{L}_\mathrm{reg}(\varphi),\\
    \mathcal W(I, \varphi) := I \circ \varphi,
    \end{gathered}
\end{equation}

\noindent
where $I_f$ and $I_m$ denote the fixed and moving images, respectively, either both 2D images with $n$ pixels, $I_f, I_m \in \R^{n}$, or both volumes with $d$ slices, $I_f, I_m \in \R^{n \times d}$. The transformation $\varphi$ belongs to the transformation space $\mathcal{T}$ (rigid, affine or deformable). The operator $\mathcal{W}(\cdot, \varphi)$ warps the moving image according to $\varphi$, and $\mathcal{S}(\cdot,\cdot)$ is a similarity measure between the fixed and warped images, to be maximised, such as normalised cross-correlation or mutual information (see Section~\ref{sec:supvsunsup}).

The transformation $\varphi$ may consist of a global component $\varphi_r$, which is rigid (3 degrees of freedom in 2D, 6 in 3D) or affine (6 and 12 parameters, respectively), and, in the deformable setting, of a dense deformation $\varphi_d$, stored on the image grid as its displacement field $\varphi_d - \mathrm{id} \in \R^{n \times 2}$ (2D) or $\R^{n \times d \times 3}$ (3D), where $\mathrm{id}$ is the identity map. The overall transformation is then the composition $\varphi = \varphi_d \circ \varphi_r$, so that the moving image is first deformed by $\varphi_d$ in its own frame and then globally transformed by $\varphi_r$. In 3D, a rigid transformation acts as $\varphi_r(x) = Rx + b$, with a translation $b \in \R^3$ and a rotation matrix $R$ in the special orthogonal group $\mathrm{SO}(3)$ of $3 \times 3$ matrices with $R^{\top} = R^{-1}$ and $\det R = 1$; these transformations form the special Euclidean group $\mathrm{SE}(3)$, to which 6-DoF poses belong. Since anatomy is neither mirrored nor folded between acquisitions, $\mathcal T$ is restricted to invertible, orientation-preserving transformations, whose Jacobian determinant $\det \nabla\varphi(x)$ is positive at every point $x$; for a rigid transformation, this is the condition $\det R = 1$. The usual rotation parametrisations, such as Euler angles, quaternions or the axis--angle representation, satisfy it by construction, whereas for dense fields it is guaranteed only by diffeomorphic parametrisations, such as an integrated velocity field, and is otherwise only encouraged by the regulariser (Section~\ref{sec:problem-formulation:regularization}). 
It is worth mentioning that parametrization can lower substantially the degrees of freedom required for both rigid and deformable registration. 

The regularization term $\mathcal{L}_{\mathrm{reg}}$ penalizes undesirable transformations, typically encouraging smoothness of the deformation field. The hyperparameter $\lambda \geq 0$ balances similarity and regularity. The optimal transformation $\varphi^*$ in Eq.~\eqref{eq:reg_general} maximises similarity while keeping the regularity penalty small.

\subsection{Cross-Dimensional Registration}

In the 2D-to-3D setting, the formulation must explicitly account for dimensional mismatch. A projection operator $\mathcal{P}$ must be introduced that allows for same-dimensionality similarity matching

\begin{equation}
\label{eq:reg_crossdim}
    \varphi^* = \argmax_{\varphi \in\mathcal{T}}\; \mathcal S\Bigl(I_f, \mathcal P\bigl(\mathcal W(V_m, \varphi)\bigr)\Bigr) - \lambda\,\mathcal{L}_\mathrm{reg}(\varphi),
\end{equation}

\noindent
where $V_m\in\R^{n\times d}$ represents the 3D volume to register to the 2D fixed image $I_f\in\R^n$. Depending on the imaging modality, $\mathcal{P}$ takes one of two forms:

\begin{itemize}
    \item \textbf{Projective}: For X-ray or fluoroscopy-based registration, $\mathcal P$ simulates radiographic image formation through ray integration, commonly implemented as DRR generation. Differentiable DRR pipelines enable gradient-based optimization and end-to-end training \cite{gaoFullyDifferentiableFramework2024, gopalakrishnan2022fast, gopalakrishnanIntraoperative2D3D2024}.
    \item \textbf{Planar}: For ultrasound or MRI registration, $\mathcal P$ extracts a 2D plane from the 3D volume at a given pose \cite{khawaledSelfattentionModelRobust2025, xuSVoRTIterativeTransformer2022, mezheritskyPopulationbased3DRespiratory2022}.
\end{itemize}

\noindent
For notational simplicity, $\mathcal P$ will denote either operation mechanism throughout this review.
Additionally, we adopt the image (voxel) reference frame: a 2D 
slice lies in the $xy$-plane, and the $z$-axis denotes the through-plane 
direction, orthogonal to the slice plane and aligned with the stacking 
direction of serial slices. In projective settings, the $z$-axis analogously 
refers to the depth direction along the projection ray. Unless otherwise 
noted, all axis references follow this convention.

\subsection{Learning-Based Registration}

In a deep learning setting, the per-case optimisation over $\varphi$ in Eq.~\eqref{eq:reg_crossdim} is replaced by a neural network $g_\theta$ with learnable parameters $\theta \in \Theta$, which maps an image--volume pair directly to a transformation:

\begin{equation}
\label{eq:learned}
    \varphi_\theta = g_\theta(I_f, V_m).
\end{equation}

\noindent
Depending on the method, $g_\theta$ outputs the global parameters $\varphi_r$, the dense deformation $\varphi_d$, or both. Once trained as described next, the network with learned parameters $\hat\theta$ returns for a new case the estimate $\hat\varphi = \varphi_{\hat\theta} \approx \varphi^*$ of the optimal transformation in Eq.~\eqref{eq:reg_crossdim}; throughout the paper, a hat denotes an estimate. 

\subsection{Supervised and Self-Supervised Learning}

In supervised learning, the optimal parameters $\theta^*$ are obtained by minimising a loss $\mathcal{L}$ between the predicted and the ground-truth transformation over a dataset of image--volume pairs with known transformations, $(I_f, V_m, \varphi_{\mathrm{GT}}) \sim \mathcal D$, where $\mathcal D$ denotes the data distribution. The learning objective is

\begin{equation}
\label{eq:supervised-weigths}
    \theta^* = \argmin_{\theta \in \Theta}\, \mathbb{E}\Bigl[ \mathcal{L}(\varphi_\theta, \varphi_\mathrm{GT}) + \lambda\,\mathcal L_\mathrm{reg}(\varphi_\theta)\Bigr],
\end{equation}

\noindent
where $\mathcal{L}$ measures the discrepancy between predicted and ground-truth transformation parameters, for example a mean squared error or a geodesic distance on $\mathrm{SO}(3)$ (see Section~\ref{sec:supvsunsup:sup}). In this setting, learning is guided by explicit supervision on the transformation parameters.
In contrast, when ground-truth transformations are most usually unavailable, learning can be formulated in a self-supervised manner. Let $(I_f, V_m) \sim \mathcal D$. The objective becomes

\begin{equation}
\label{eq:ssl-weights}
    \begin{aligned}
        \theta^* = \argmax_{\theta\in\Theta}\, \mathbb{E}\Bigl[&\mathcal S\bigl(I_f, \mathcal P(\mathcal W(V_m, \varphi_\theta))\bigr) \\
        &- \lambda\,\mathcal L_\mathrm{reg}(\varphi_\theta)\Bigr].
    \end{aligned}
\end{equation}

Here, the model parameters are optimised such that the predicted transformation maximises the similarity between the fixed image and the projection of the warped volume, i.e. the same similarity $\mathcal S$ as in Eq.~\eqref{eq:reg_crossdim} now acts as the training signal. 

\subsection{Regularization Strategies}
\label{sec:problem-formulation:regularization}

The most common form of regularization is the diffusion regulariser, which penalizes squared $\mathcal{L}_2$-norm of the spatial gradients of the displacement fields for deformable registration. This discourages large differences between neighbouring displacements \cite{chenSurveyDeepLearning2025}. An alternative is the bending energy regulariser, which penalizes second-order derivatives and has the property of being invariant to affine transformations, potentially eliminating the need for a separate pre-alignment step \cite{rueckert2002nonrigid}. To constrain rigid bodies such as bones, Moving Least Squares (MLS) can be employed, constraining the shape changes to be as rigid as possible \cite{khawaledSelfattentionModelRobust2025}. Alternatively, a diffusion or bending energy regulariser can be employed and modulate its strength spatially, with stronger weights on bone regions \cite{chen2023spatially}. Jacobian-based penalties offer an additional mechanism by directly penalizing non-positive Jacobian determinants, which indicate locally non-invertible (folded) transformations \cite{chenSurveyDeepLearning2025}.

Beyond explicit regularization, transformations can also be implicitly regularized through architectural choices or invertibility constraints. In Implicit Neural Representations (INRs), for instance, the activation function of the network determines how smooth the represented deformation is (Section~\ref{sec:arch-inr}).  In the reviewed literature, regularization remains underspecified. Table~\ref{tab:loss-functions-summary} shows that only ten studies include an explicit regularization term, eight of them among the thirteen deformable ones; most of the other deformable methods constrain the field through a low-dimensional parametrisation, such as PCA modes or a statistical shape model, or by integrating a velocity field \cite{zhangDeformableRegistrationLateral2021, khamenehAutomaticCNNbased3D2025, mhiriNeuralPatientspecific3D2D2025b, hindleyOpenSourceDeep2025}. Where a regulariser is used, its choice and its weight $\lambda$ are rarely justified or systematically compared.

\subsection{From the Formulation to the Structure of the Review}
\label{sec:problem-formulation:roadmap}

The learning formulation of Eqs.~\eqref{eq:learned}--\eqref{eq:ssl-weights} separates three design decisions that every learning-based method makes, and Sections~\ref{sec:supvsunsup} to~\ref{sec:architectures} examine the included studies from the perspective of each decision in turn. Section~\ref{sec:supvsunsup} concerns the training signal, that is, whether $\theta$ is learned from ground-truth transformations as in Eq.~\eqref{eq:supervised-weigths}, from image similarity $\mathcal S$ as in Eq.~\eqref{eq:ssl-weights}, or from both. Sections~\ref{sec:direct} and~\ref{sec:iterative} concern inference: whether $g_\theta$ produces the transformation in a single forward pass or the estimate is refined over several steps, by repeated network passes or by optimisation at test time. Section~\ref{sec:architectures} concerns the parametrisation of $g_\theta$, namely the deep learning backbones that encode $I_f$ and $V_m$, the differentiable geometric operators that implement $\mathcal P$ and $\mathcal W$ inside the network, and generative and implicit representations. The two sections that follow address the step towards clinical use: Section~\ref{sec:eval} reviews how the estimated transformation is evaluated, and Section~\ref{sec:domain-adaptation} how the studies handle domain shift, whether between simulated and clinical images or between imaging modalities.

\section{Learning Paradigms}
\label{sec:supvsunsup}
Deep Learning-based registration methods can be divided into supervised and self-supervised approaches. Supervised learning requires ground-truth labels representing the exact spatial relationship between the 2D image and the 3D volume, while self-supervised learning utilizes image similarity metrics to drive alignment. Several studies describe training on simulated transformations as self-supervised, because the labels are generated automatically rather than annotated \cite{gopalakrishnanIntraoperative2D3D2024, gopalakrishnanRapidPatientspecificNeural2025, liDiffusionstimulatedCTUSRegistration, oya2D3DDeformable2024}. We classify studies by their training signal: a loss on a known transformation counts as supervision, and its combination with an image-similarity term as a hybrid approach (Table~\ref{tab:loss-functions-summary}); the one study trained by reinforcement learning is labelled separately (Section~\ref{sec:supvsunsup:rl}). 

\subsection{Supervised Learning}
\label{sec:supvsunsup:sup}
Supervised methods, adopted by 27 of the 41 included studies, train $g_\theta$ by minimising a loss between the predicted and the ground-truth transformation, as in Eq.~\eqref{eq:supervised-weigths}. The ground truth is usually known by construction: a transformation is sampled and applied to the preoperative volume, and the training image is rendered or sliced from the transformed volume, so that large labelled training sets can be generated without annotation \cite{miaoCNNRegressionApproach2016, gopalakrishnanIntraoperative2D3D2024}. Of the 27 supervised studies, 22 train on synthetic data only (Table~\ref{tab:data-extraction-summary}). A loss on the transformation parameters gives a direct error signal, which offers faster and more stable convergence than an image-based similarity.

The weakness of supervised training lies in the labels of real images: exact ground-truth transformations between a clinical 2D image and a 3D volume are rarely available, and those that can be obtained often come from another registration method or from manual alignment. Random errors in such labels largely average out over a large training set, but systematic ones do not, and the network learns them. This limitation is particularly evident in the case of deformable registration, where ground truth deformation fields are frequently generated using alternative deformable registration algorithms such as Elastix \cite{shamonin2014fast, elastix} and SyN \cite{avants2011reproducible} rather than direct measurement \cite{hindleyOpenSourceDeep2025, zhangDeformableRegistrationLateral2021, jiangSemanticallyConsistentDeformable2D3D2025a}. In rigid registration, clinical annotation is technically feasible but difficult, which keeps annotated clinical data scarce and explains the reliance on synthetic transformations. Synthetic training data, however, introduce a domain gap (Section~\ref{sec:domain-adaptation}), and performance is ultimately bounded by the quality of the labels.

The loss functions used in supervised models are varied.
For both rigid and deformable registration, Table \ref{tab:loss-functions-summary} shows that the most common baseline is a standard Mean Squared Error (MSE) or Mean Absolute Error (MAE), targeting the distance of the predicted transformation parameters \cite{miaoCNNRegressionApproach2016, khamenehAutomaticCNNbased3D2025, guoUltrasoundFrametoVolumeRegistration2023a}. Specifically for rigid registration, while these are effective for translational coordinates, they often struggle with the non-linearity of 3D rotations. To address this, another group of studies \cite{mohsenisalehiRealTimeDeepPose2019, downsImprovingAutomaticCerebral2025, wuDVAPRegDualviewAnatomical2026} incorporate a geodesic loss, the angle of the relative rotation between prediction and ground truth on the rotation group $\mathrm{SO}(3)$ defined in Section~\ref{sec:problem-formulation:classical}:
\begin{equation}
\label{eq:geodesic}
    \mathcal L_\mathrm{geo}(\hat R, R_\mathrm{GT}) = \arccos\biggl(\frac{\operatorname{tr}\bigl(\hat R^{\top} R_\mathrm{GT}\bigr) - 1}{2}\biggr),
\end{equation}
where $\hat R, R_\mathrm{GT} \in \mathrm{SO}(3)$ are the predicted and ground-truth rotation matrices. Unlike an $L_1$ or $L_2$ loss on Euler angles, it measures the rotation error itself and does not depend on how rotations are parametrised. These parametrization choices are further discussed in Section \ref{sec:eval:parameter-wise}. 

These losses are often complemented by additional anatomical constraints and regularization terms. For example, \citet{quTransformerbased2D3D2024b} introduced a landmark-based loss, \citet{oya2D3DDeformable2024} used a vertex position loss on a deformable mesh model, while \citet{lecomte-denisDomainAgnostic2D3D2025} computed MSE on projected displacement fields and Dice loss on projected segmentation masks. \citet{facenteRealtimeAlignment3D2026} instead train only on the projected distance of CT landmarks, extracted automatically from a surface mesh of the CT.
\subsection{Self-Supervised Learning}
\label{sec:supvsunsup:unsup}
Self-supervised networks adopt the same intensity-based optimisation principle as traditional registration methods, which do not use deep learning and optimise the transformation separately for each case as in Eq.~\eqref{eq:reg_crossdim}: a 2D image is synthesised from the 3D volume with differentiable geometric operators, and the discrepancy between the synthesised and the fixed image drives the registration.
 
 The main advantage of this paradigm is its independence from ground-truth labels. This allows the direct integration of clinical data during training. This is because the similarity between the transformed and original volume serves as the sole training signal. Nevertheless, image similarity objectives are non-convex and can contain multiple local minima. In mono-modal settings, MSE is often used as a similarity measure \cite{zhangUnsupervised2D3DDeformable2021, yeNonorthogonalKVImaging2025, yiasemis2024deep}. Yet, image similarity alone does not guarantee anatomically correct alignment (see Sections \ref{sec:eval:geometrical}, \ref{sec:eval:image-similarity}).%

Loss functions employed operate in the image domain. The Normalized Cross Correlation Coefficient (NCC) is the most common metric for X-ray-to-CT tasks due to its robustness to linear intensity variations \cite{gopalakrishnanIntraoperative2D3D2024, gopalakrishnanRapidPatientspecificNeural2025, wu3DIsotropicHighResolution2025, jiangSemanticallyConsistentDeformable2D3D2025a, tianLiftRegLimitedAngle2022}. For modalities where intensity relationships are non-linear, structural or gradient-based metrics are preferred: \citet{yeNonorthogonalKVImaging2025} combine intensity and gradient terms to prioritize anatomical contours over raw pixel values, and \citet{zhangEnhancingCNNRegressors2026} compare image gradients in the self-supervised stage of their hybrid training. The Structural Similarity Index Measure (SSIM) \cite{wang2004ssim}, which compares local luminance, contrast and structure, is not part of the final training objective of any reviewed study; \citet{dupuy2D3DDeep2023a} tested an SSIM penalty alongside their parameter loss for ultrasound and found that it brought little improvement. Mutual Information (MI), the classical multi-modal similarity measure, assumes only a statistical dependence between the intensities of the two images rather than a linear relationship, which makes it robust to differences in appearance between modalities \cite{maes1997mi}; it does not, however, compensate for anatomical change or for structures visible in only one image \cite{pluim2003mi}. \citet{chiTwostepDeepLearning2023} instead compare modality-independent neighbourhood descriptors (MIND) of the warped CT and the US, together with a soft Dice between kidney probability maps, and train their network entirely without ground truth, first on a population of generated CT--US pairs and then on the first respiratory cycle of each patient. \citet{zheng2022unsupervised} employed MI as a content-preservation loss during domain adaptation to ensure that pixel-level style transfer does not alter the underlying anatomical structure. As in the supervised case, the similarity can be complemented by anatomical guidance: \citet{jiangSemanticallyConsistentDeformable2D3D2025a} incorporated a semantic consistency loss based on mandibular similarity to anchor the registration.

\subsection{Hybrid approaches}
\label{sec:supvsunsup:hybrid}
The distinction between supervised and self-supervised paradigms reflects a trade-off between optimisation stability and clinical realism. While supervised methods offer stable convergence, they are limited by label scarcity and the resulting reality gap. Conversely, self-supervised methods bypass the need for labels, but often require larger amounts of data to converge.

Hybrid approaches combine the two signals to retain the advantages of both. Most do so jointly, adding an image-similarity term to a loss on the simulated pose: \citet{gopalakrishnanIntraoperative2D3D2024, gopalakrishnanRapidPatientspecificNeural2025} combine, in DiffPose and xvr, a geodesic pose loss with multiscale NCC between the X-ray and the DRR rendered at the predicted pose, \citet{liDiffusionstimulatedCTUSRegistration} a pose loss with an MSE between the corresponding CT slices, and \citet{zhangUltrasoundSlicetoVolumeRegistration2026} a pose loss with an intensity MSE computed only on artifact-free regions. \citet{mezheritskyPopulationbased3DRespiratory2022} pair an image MSE with a loss on the latent codes of deformations estimated by a pretrained registration network. \citet{zhangEnhancingCNNRegressors2026} instead use two stages on simulated data, a supervised pre-training followed by a fine-tuning that adds image-gradient and feature similarity terms to the pose loss. This order reverses the usual one in deep learning, where self-supervised pretraining on unlabelled data precedes supervised fine-tuning on a small labelled set \cite{chen2020simclr, azizi2021big}. In registration, labels are available for synthetic data, whereas clinical images can only be used through the self-supervised objective, so the supervised stage comes first.

\subsection{Reinforcement Learning}
\label{sec:supvsunsup:rl}
In reinforcement learning (RL), an agent learns by trial and error which action to take in a given state: after each action it receives a scalar reward, and it is trained to maximise the reward accumulated over a sequence of actions rather than to reproduce a target output \cite{sutton2018reinforcement}. \citet{zhangWarmstartedReinforcementLearning2026} train their network as such an agent with proximal policy optimisation (PPO): the reward after every action is the reduction in mesh-vertex error with respect to the ground-truth pose, so the ground truth is used only indirectly, and the agent learns a sequence of corrections and when to stop. Table~\ref{tab:loss-functions-summary} therefore lists the study under its own label (RL); at inference, the agent registers iteratively (Section~\ref{sec:multipass}).

\begin{table*}[!htbp]
\centering
\caption{Categorization of Loss Functions and Supervision across the Reviewed Articles.}
\label{tab:loss-functions-summary}
\footnotesize %
\begin{tabularx}{\textwidth}{@{\extracolsep{\fill}}l c ccccccc}
\toprule
\textbf{Article} & 
\textbf{\makecell{Sup. \\ Type}} & 
\textbf{\makecell{$L_1$/ \\ $L_2$}} & 
\textbf{\makecell{Geo- \\ desic}} & 
\textbf{NCC} & 
\textbf{MI} &
\textbf{\makecell{Anat.}} & 
\textbf{\makecell{Reg.}} & 
\textbf{Other} \\
\midrule
\citet{caoNonrigidImagevolumeRegistration2025} & Sup. & \checkmark &  &  &  & \checkmark & \checkmark &  \\
\citet{chiTwostepDeepLearning2023} & SSL &  &  &  &  & \checkmark & \checkmark & \checkmark \\
\citet{daiVolumetricTumorTracking2024} & Sup. & \checkmark &  &  &  & \checkmark & \checkmark &  \\
\citet{downsImprovingAutomaticCerebral2025} & Sup. &  & \checkmark & \checkmark & \checkmark & \checkmark &  &  \\
\citet{dupuy2D3DDeep2023a} & Sup. & \checkmark &  &  &  &  &  &  \\
\citet{facenteRealtimeAlignment3D2026} & Sup. & \checkmark &  &  &  & \checkmark &  &  \\
\citet{gaoFullyDifferentiableFramework2024} & Sup. &  & \checkmark & \checkmark &  &  &  & \checkmark \\
\citet{gopalakrishnanIntraoperative2D3D2024} & Hyb. &  & \checkmark & \checkmark &  &  &  &  \\
\citet{gopalakrishnanRapidPatientspecificNeural2025} & Hyb. &  & \checkmark & \checkmark &  &  &  &  \\
\citet{guoUltrasoundFrametoVolumeRegistration2023a} & Sup. & \checkmark &  &  &  &  &  & \checkmark \\
\citet{hindleyOpenSourceDeep2025} & Sup. & \checkmark &  &  &  &  &  &  \\
\citet{jiangSemanticallyConsistentDeformable2D3D2025a} & SSL &  &  & \checkmark &  & \checkmark & \checkmark &  \\
\citet{khamenehAutomaticCNNbased3D2025} & Sup. & \checkmark &  &  &  &  &  &  \\
\citet{khawaledSelfattentionModelRobust2025} & Sup. & \checkmark &  &  &  &  &  &  \\
\citet{kobayashiShapeRegistrationLaparoscopic2025} & Sup. &  &  &  &  &  &  & \checkmark \\
\citet{lecomte-denisDomainAgnostic2D3D2025} & Sup. & \checkmark &  &  &  & \checkmark &  &  \\
\citet{liDiffusionstimulatedCTUSRegistration} & Hyb. & \checkmark &  &  &  &  &  & \checkmark \\
\citet{mezheritskyPopulationbased3DRespiratory2022} & Hyb. & \checkmark &  &  &  &  & \checkmark &  \\
\citet{mhiriNeuralPatientspecific3D2D2025b} & Sup. & \checkmark &  &  &  & \checkmark &  &  \\
\citet{miaoCNNRegressionApproach2016} & Sup. & \checkmark &  &  &  &  &  &  \\
\citet{mohsenisalehiRealTimeDeepPose2019} & Sup. & \checkmark & \checkmark &  &  &  &  &  \\
\citet{oya2D3DDeformable2024} & Sup. & \checkmark &  &  &  & \checkmark & \checkmark &  \\
\citet{qianHybridPopulationbasedPatientspecific2025} & SSL & \checkmark &  &  &  &  & \checkmark &  \\
\citet{quTransformerbased2D3D2024b} & Sup. & \checkmark &  &  &  & \checkmark &  &  \\
\citet{soormallyDeepSensorlessTracking2025} & Sup. & \checkmark &  &  &  &  &  & \checkmark \\
\citet{tianLiftRegLimitedAngle2022} & SSL &  &  & \checkmark &  &  & \checkmark &  \\
\citet{wangContrastiveLearningBased2024} & Sup. & \checkmark &  &  &  &  &  & \checkmark \\
\citet{wangPoseawareDeepPerceptual2026} & Sup. &  &  &  &  &  &  & \checkmark \\
\citet{weiDeepLearningApproach2021} & Sup. &  &  &  &  & \checkmark &  &  \\
\citet{wu3DIsotropicHighResolution2025} & SSL & \checkmark &  & \checkmark &  &  &  &  \\
\citet{wuDVAPRegDualviewAnatomical2026} & Sup. & \checkmark & \checkmark &  &  & \checkmark & \checkmark & \checkmark \\
\citet{xuSVoRTIterativeTransformer2022} & Sup. & \checkmark &  &  &  &  &  &  \\
\citet{yeNonorthogonalKVImaging2025} & SSL & \checkmark &  &  &  &  &  & \checkmark \\
\citet{yeRobustMethodReal2025} & Sup. & \checkmark &  &  &  &  &  &  \\
\citet{zhangUnsupervised2D3DDeformable2021} & SSL & \checkmark &  &  &  &  & \checkmark &  \\
\citet{zhangDeformableRegistrationLateral2021} & Sup. & \checkmark &  &  &  &  &  &  \\
\citet{zhangEnhancingCNNRegressors2026} & Hyb. & \checkmark &  &  &  &  &  & \checkmark \\
\citet{zhangUltrasoundSlicetoVolumeRegistration2026} & Hyb. & \checkmark &  &  &  &  &  &  \\
\citet{zhangWarmstartedReinforcementLearning2026} & RL &  &  &  &  &  &  & \checkmark \\
\citet{zhengPairwiseDomainAdaptation2018} & Sup. & \checkmark &  &  &  &  &  & \checkmark \\
\citet{zheng2022unsupervised} & Sup. & \checkmark &  &  & \checkmark &  &  & \checkmark \\
\bottomrule
\end{tabularx}
\vspace{2pt}
\begin{flushleft}
\scriptsize
\textbf{Legend:} \textbf{Sup.}: Supervised; \textbf{SSL}: Self-Supervised; \textbf{Hyb.}: Hybrid; \textbf{RL}: Reinforcement Learning;
\textbf{$L_1$/$L_2$}: mean absolute or squared error on transformation parameters, points or intensities; \textbf{MI}: Mutual Information; \textbf{Anat.}: Anatomical/Semantic; \textbf{Reg.}: Regularization.
\end{flushleft}
\end{table*}

\section{Direct Regression}
\label{sec:direct}
\begin{figure*}[!t]
    \centering
    \includegraphics[width=0.85\textwidth]{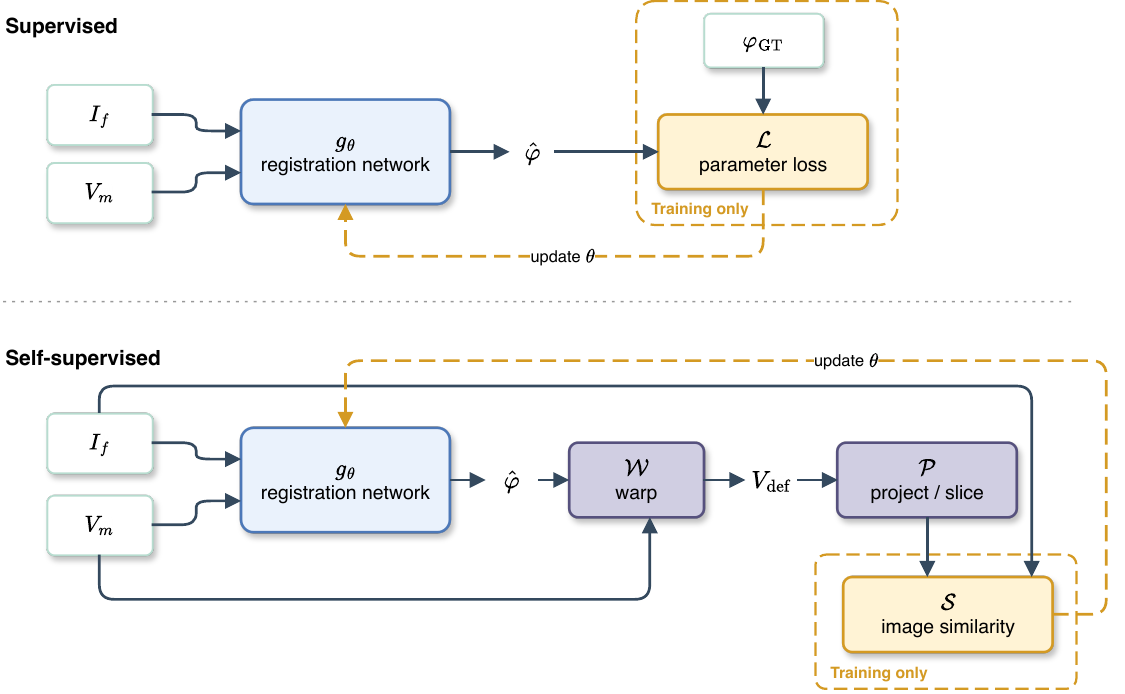}
    \caption{Direct regression. Top: supervised training as in Eq.~\eqref{eq:supervised-weigths}, where the predicted transformation $\hat\varphi$ is compared with a ground truth $\varphi_\mathrm{GT}$. Bottom: self-supervised training as in Eq.~\eqref{eq:ssl-weights}, where the deformed volume $V_\mathrm{def} = \mathcal W(V_m, \hat\varphi)$ is projected or sliced and compared with $I_f$. Colour code, used in all schematics of this review: blue, learnable network; green, regression head; purple, differentiable geometric operator; amber, similarity measure or loss; red, classical optimiser; white, image, volume or label map. Dashed amber arrows denote weight updates (training only), dashed purple arrows iteration.}
    \label{fig:direct}
\end{figure*}
Despite the progress in the field, the core learning-based registration pipeline, both in its training and inference setup, has remained fundamentally unchanged over the years. Figure  \ref{fig:direct} illustrates the direct regression paradigm, in which a general transformation $\hat\varphi$ is inferred in a single forward pass for both supervised and self-supervised approaches. The general pipeline takes  $I_f$ and $V_m$ as inputs and feeds them into a regressor, typically a neural network. 
For the supervised case, the predicted transformation $\hat\varphi$ is compared with a ground-truth $\varphi_\mathrm{GT}$. In the self-supervised case, conversely, $\hat\varphi$ is applied to $V_m$, obtaining  $V_\mathrm{def}$. Finally, a slice or 2D projection is computed and compared using a similarity metric to $I_f$.

Direct regression is the most common pipeline among the reviewed articles, adopted by 23 of the 41 studies, while the other 18 refine their estimate iteratively (Section~\ref{sec:iterative}). Its appeal lies in speed: a single forward pass of a neural network often takes only milliseconds, as real-time guidance requires \cite{mezheritskyPopulationbased3DRespiratory2022, chiTwostepDeepLearning2023, quTransformerbased2D3D2024b,dupuy2D3DDeep2023a, soormallyDeepSensorlessTracking2025, jiangSemanticallyConsistentDeformable2D3D2025a}. Of these, 14 address rigid and 9 deformable registration.

The speed of this single-step approach may, however, come at the expense of  accuracy, which is a central limitation of this approach, particularly under large misalignments, exhibiting a limited capture range compared to iterative approaches (see Section \ref{sec:iterative}).

Works included in this review aim to tackle this speed-accuracy trade-off in several ways. Self-attention mechanisms have been introduced to generate pixel-wise relevance scores, allowing the network to suppress  noise while focusing on informative anatomical regions \cite{khawaledSelfattentionModelRobust2025} (Section~\ref{sec:architectures:attention}). Several studies implement a patient-specific training approach, utilizing exclusively synthetic data derived from the imaging of the same patient, or from  a population model, followed by a final fine-tuning step on patient-specific data \cite{mhiriNeuralPatientspecific3D2D2025b, liDiffusionstimulatedCTUSRegistration, chiTwostepDeepLearning2023}. 

\citet{wuDVAPRegDualviewAnatomical2026} change both inputs of $g_\theta$ in Eq.~\eqref{eq:learned}. For per-vertebra navigation in spinal surgery, the fixed image is a pair of anteroposterior and lateral X-rays, and the moving volume is replaced by the surface mesh of the segmented vertebra, encoded by a graph convolutional network without using CT intensities. From each view, a CNN predicts intermediate geometric maps, including dense 2D--3D correspondences, which a learned PnP head fuses with the mesh features into a single 6-DoF pose. On real intraoperative X-rays, with a reference pose aligned manually by one physician, the errors are 2.81\textdegree{} and 1.82\,mm at 63\,ms per vertebra, which makes the method suitable for real-time navigation.

\section{Iterative Approaches}
\label{sec:iterative}

\begin{figure*}[tb]
    \centering
    \includegraphics[width=0.8\textwidth]{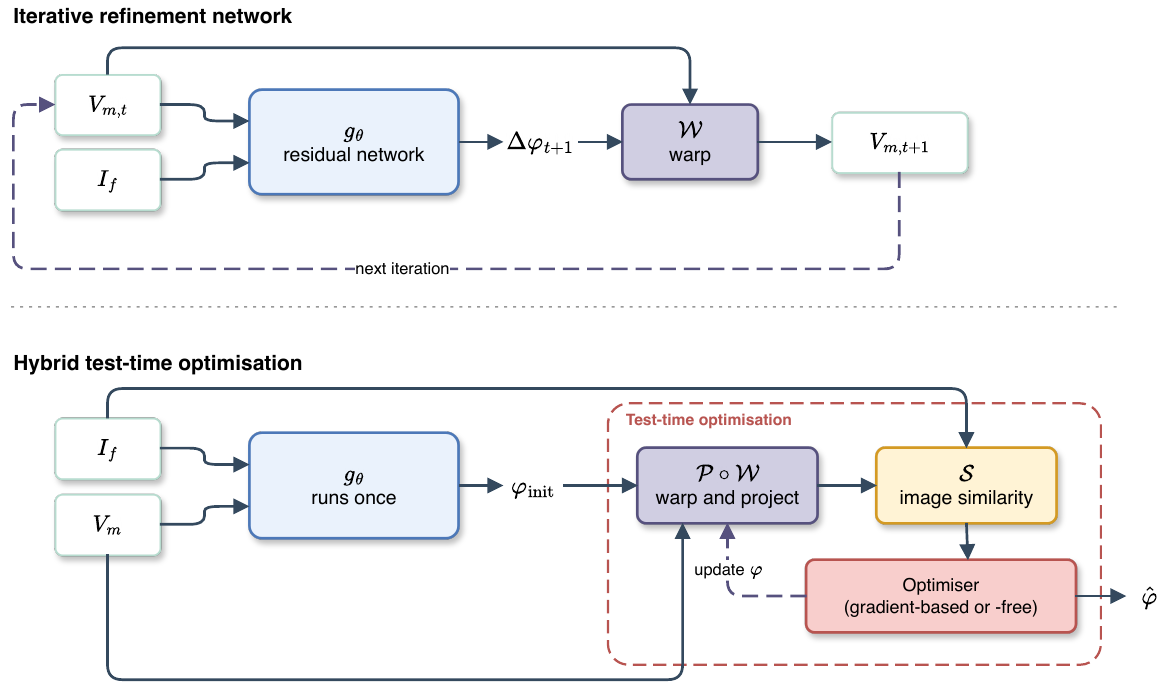}
    \caption{Iterative registration at inference. Top: iterative refinement network (Section~\ref{sec:multipass}). At each iteration, the network predicts a residual correction $\Delta\varphi_{t+1}$ from $I_f$ and the moving volume $V_{m,t}$ warped with the current estimate, and the warp applies it to give $V_{m,t+1}$, the input of the next iteration. Bottom: hybrid test-time optimisation, where a single network pass provides the initialisation $\varphi_\mathrm{init}$ for a classical optimiser, which returns the estimate $\hat\varphi$ (Section~\ref{sec:hybrid-tto}).}
    \label{fig:iterative}
\end{figure*}
In contrast to direct regression, which predicts the transformation in a single forward pass, iterative methods refine an initial estimate through successive updates. Instead of solving the full alignment problem at once, iterative registration pipelines subdivide the task of regressing the correct pose or deformation of $V_m$ into a simpler correction step that can be repeated until a stopping criterion is met. Figure~\ref{fig:iterative} contrasts the two families surveyed in this section. In iterative refinement networks (top), the network receives $I_f$ together with the moving volume $V_{m,t}$ warped with the current estimate and predicts a residual correction $\Delta\varphi_{t+1}$, which a spatial transformer applies to produce $V_{m,t+1}$ for the next iteration. In hybrid test-time optimisation (bottom), the network is run only once to obtain an initialisation $\varphi_\mathrm{init}$, and the remaining iterations are carried out by a classical optimisation-based registration through a differentiable forward model. 

Compared with direct inference, an iterative step has the main advantage of having a better initialisation. They learn from residual misalignments rather than from the full displacement error, allowing higher precision. This, however, introduces a clear speed-accuracy trade-off. Since 2D–to-3D registration methods are employed primarily for intraoperative guidance, these approaches must meet clear inference time requirements and, therefore, require careful optimisation \cite{miaoCNNRegressionApproach2016, guoUltrasoundFrametoVolumeRegistration2023a}. On the other hand, they excel in applications that do not have strict time requirements, such as Slice-to-Volume Registration (SVR) \cite{xuSVoRTIterativeTransformer2022}.

There are multiple ways of designing an iterative pipeline for registration: Iterative refinement networks embed the neural network forward pass within the iterative cycle, enabling learning from residuals. Hybrid test-time optimisation refines an initial pose predicted by a neural network with a classical optimiser, either gradient-based or gradient-free. Finally, emerging paradigms such as per-instance network fitting and diffusion-based strategies introduce alternative forms of iterative refinement. We survey each of these threads below.

\subsection{Iterative refinement networks}
\label{sec:multipass}
These methods embed the network forward pass in the iteration loop (Figure~\ref{fig:iterative}, top). Starting from an initial estimate $\varphi_0$, each iteration predicts a residual correction from the moving volume warped with the current estimate,
\begin{equation}
\label{eq:multipass}
\begin{aligned}
    V_{m,t} &= \mathcal W(V_m, \varphi_t), \\
    \Delta\varphi_{t+1} &= g_\theta(I_f, V_{m,t}), \\
    \varphi_{t+1} &= \varphi_t \circ \Delta\varphi_{t+1}, \qquad t = 0, \dots, K-1, \\
    \hat\varphi &= \varphi_K,
\end{aligned}
\end{equation}
where $K$ is the number of iterations; in projective settings, the network sees $V_{m,t}$ through its projection $\mathcal P(V_{m,t})$. By the definition of $\mathcal W$ in Eq.~\eqref{eq:reg_general}, $V_{m,t+1}$ is equally obtained by warping $V_{m,t}$ with the correction alone, $V_{m,t+1} = \mathcal W(V_{m,t}, \Delta\varphi_{t+1})$, as drawn in Figure~\ref{fig:iterative}. Each iteration only has to correct the misalignment left by the previous one, so the network operates on progressively smaller errors. $K$ is either fixed or set by a stopping criterion, such as the similarity $\mathcal S$ no longer improving. In the supervised case, $g_\theta$ is typically trained to predict the residual $\varphi_t^{-1} \circ \varphi_{\mathrm{GT}}$ from estimates $\varphi_t$ sampled around the ground truth. Here $\varphi_t^{-1}$ is the inverse of the current estimate, the transformation that undoes $\varphi_t$, so the residual is the correction that, composed with $\varphi_t$ as in Eq.~\eqref{eq:multipass}, gives exactly $\varphi_\mathrm{GT}$. Iterative refinement networks are closely related to algorithm unrolling, in which a fixed number of iterations of an optimisation algorithm are written as successive network stages whose updates are learned from data \cite{mongaAlgorithmUnrollingInterpretable2021}, so that each network pass takes the place of one iteration of a classical solver of Eq.~\eqref{eq:reg_crossdim}. When all passes are trained jointly, with a loss on every iteration, as in SVoRT \cite{xuSVoRTIterativeTransformer2022}, the pipeline is an unrolled optimisation in this sense.

\citet{miaoCNNRegressionApproach2016} pioneered this approach by decomposing 6-DoF regression into hierarchical parameter groups of increasing difficulty, each handled by a dedicated CNN stage. The cascade is repeated $K$ times, with each stage to reduce residual misalignment and progressively simplify the regression landscape for each sub-network. This enables the system to handle larger initial misalignments than a single-shot regressor. \citet{khamenehAutomaticCNNbased3D2025} follow a similar hierarchy to fit a statistical femur model to radiographs: a fixed cascade of CNNs estimates pose, shape and scale in turn, each stage working on projections of the previous stage's estimate. %

\citet{zhangDeformableRegistrationLateral2021} extended learned iteration to the deformable setting with a closed-loop feedback mechanism for lateral cephalogram to CBCT registration. Their CNN predicts a DVF parametrized by Principal Component Analysis (PCA) basis functions, which is then applied to warp the reference CBCT volume. At each iteration, a new DRR is regenerated from the newly deformed volume, allowing the network to observe the residual misalignment rather than the full initial displacement.

\citet{facenteRealtimeAlignment3D2026} use the same feedback in LXPose, a patient-specific two-stage cascade for rigid X-ray to CT registration. A first CNN regresses an initial pose from the X-ray alone; a second CNN receives the X-ray with a DRR rendered at that pose and predicts a residual correction, and can be applied for any number of passes. This learned refinement replaces the test-time optimisation of DiffPose and its successor xvr \cite{gopalakrishnanIntraoperative2D3D2024, gopalakrishnanRapidPatientspecificNeural2025} (Section~\ref{sec:hybrid-tto}). On the DeepFluoro dataset, a single pass takes 20\,ms and twelve passes 187\,ms, against 3.6\,s for DiffPose and 7.8\,s for xvr. In the authors' comparison, with a shared backbone and training setup, its projection error is within 0.4\,mm of xvr's after one pass and within 0.15\,mm after twelve, but its 3D error is considerably larger, mostly in depth.

\citet{zhangWarmstartedReinforcementLearning2026} follow the same loop with a reinforcement-learning agent, for rigid registration of a CT-derived liver model to laparoscopic video. Each step is a single discrete move: the agent chooses which degree of freedom to adjust, in which direction and by how much, or decides to stop, and both the moves and the stopping point are learned with proximal policy optimisation (Section~\ref{sec:supvsunsup:rl}). Registration therefore takes many small steps, 40 to 256 or about 1 to 5\,s. On the public four-patient benchmark also used by \citet{mhiriNeuralPatientspecific3D2D2025b}, the mean tumour TRE is 15.70\,mm, against 24.83\,mm for Mhiri et al. in the authors' comparison.

\citet{guoUltrasoundFrametoVolumeRegistration2023a} use the same loop for ultrasound frame-to-volume registration, re-slicing the volume at the current estimate and correcting it until the similarity between slice and frame stops improving; each frame starts from the estimate of the previous one. \citet{zhengPairwiseDomainAdaptation2018} evaluate their domain adaptation module (Section~\ref{sec:domain-adaptation}) inside a similar loop, with a reinforcement-learning agent that adjusts the pose from the X-ray and a DRR rendered at the current estimate.

In fetal brain MRI, severe and unpredictable motion prevents the acquisition of high-quality 3D volumes, hence the reference volume must itself be reconstructed from the acquired 2D slices, making SVR and volume reconstruction mutually dependent. \citet{xuSVoRTIterativeTransformer2022} addresses this with the Slice-to-Volume Registration Transformer (SVoRT), which performs $K$ iterations, each consisting of two connected steps: a Slice-Volume Transformer (SVT) that regresses residual transformations, and a volume estimation step that reconstructs an updated 3D volume from the refined slice positions. What distinguishes this from earlier methods is the mutual refinement: the registration target itself improves across iterations, as better-aligned slices produce a sharper reconstructed volume, which in turn provides better reference features for the next registration step. \citet{wu3DIsotropicHighResolution2025} alternate in the same way between a slice-to-volume network and a super-resolution reconstruction network, both trained for the individual scan without external data. This coupling is specific to settings where the 3D reference is itself being estimated, and it does not transfer to scenarios with a fixed, pre-acquired reference volume.

The accuracy of iterative refinement networks, and in general of every neural network, is ultimately bound by what the network has learned during training. This limitation motivates the next class of approaches, where test-time optimization through physics-based rendering breaks free from purely learned constraints.

\subsection{Hybrid Test-time optimisation via Differentiable Geometric Operators}
\label{sec:hybrid-tto}
Hybrid test-time optimisation runs the network only once, to initialise the transformation (Figure~\ref{fig:iterative}, bottom), whereas in iterative refinement networks (Section~\ref{sec:multipass}) the network computes every update. Its estimate starts a classical solution of Eq.~\eqref{eq:reg_crossdim}, in which the network weights are fixed and only the transformation is optimised for the case at hand, by gradient-based optimisation through a differentiable forward model (Section~\ref{sec:architectures:differentiable-geometric-operators}):
\begin{equation}
\label{eq:tto}
\begin{aligned}
    \varphi^{(0)} &= \varphi_\mathrm{init} = g_\theta(I_f, V_m), \\
    \Delta\varphi^{(j+1)} &= U\bigl(\nabla_\varphi J(\varphi^{(j)})\bigr), \\
    \varphi^{(j+1)} &= \varphi^{(j)} \circ \Delta\varphi^{(j+1)},
\end{aligned}
\end{equation}
where $J(\varphi) = \mathcal S\bigl(I_f, \mathcal P(\mathcal W(V_m,\varphi))\bigr) - \lambda\,\mathcal L_\mathrm{reg}(\varphi)$ is the objective of Eq.~\eqref{eq:reg_crossdim} and $U$ is the update rule of the test-time optimiser, which turns the gradient of $J$ with respect to $\varphi$ into a small correction $\Delta\varphi^{(j+1)}$; the last iteration gives the estimate $\hat\varphi$. This optimiser acts only on $\varphi$: the network weights stay fixed at their learned values $\hat\theta$, so $U$ is distinct from the optimiser that learned $\theta$ in Eqs.~\eqref{eq:supervised-weigths}--\eqref{eq:ssl-weights}, even when both use the same algorithm. The update has the same form as Eq.~\eqref{eq:multipass}; what changes is that the correction is computed by the optimiser instead of the network. When $\varphi$ is represented by a Euclidean parameter vector, the composition is replaced by adding the step to the parameters, and plain gradient ascent scales the gradient by a single step size. Rotation and translation parameters, however, differ in units and scale, which a single step size handles poorly. \citet{gopalakrishnanIntraoperative2D3D2024} (DiffPose) and \citet{downsImprovingAutomaticCerebral2025} use Adam, which adapts the step of each parameter, and DiffPose additionally sets learning rates three orders of magnitude apart for the rotational and translational components. For rigid poses, DiffPose also parametrises the pose in the Lie algebra $\mathfrak{se}(3)$ of the group $\mathrm{SE}(3)$ (Section~\ref{sec:problem-formulation:classical}), that is, in the tangent space of the group at the identity. Its elements are twists $\xi = (\omega, v) \in \R^6$, which combine a rotation vector $\omega$, the rotation axis scaled by the angle, with a translational component $v$, and the exponential map $\exp\colon \mathfrak{se}(3) \to \mathrm{SE}(3)$ turns a twist into a rigid transformation. DiffPose writes the pose as $\exp(\xi)$ and takes gradient steps on $\xi$, which treats rotation and translation jointly and avoids the singularities of Euler angles. In a systematic comparison in the same study, this parametrisation gave higher sub-millimetre success rates and lower mTRE than parametrisations that treat rotation and translation separately, such as Euler angles, quaternions or higher-dimensional rotation representations, although the evidence comes from a single study. The network supplies a starting point within the capture range, and the optimisation can refine it beyond the accuracy that the network alone achieves. %

\citet{gopalakrishnanIntraoperative2D3D2024}
 established the foundational framework for this approach with DiffPose. In a preoperative phase, a patient-specific CNN is trained exclusively on synthetic X-rays rendered from the patient's own CT under random rigid perturbations, requiring no manual annotations. At inference, the CNN provides an initial pose estimate that is subsequently refined through iterative gradient-based optimisation, with the differentiable renderer DiffDRR (Section~\ref{sec:architectures:differentiable-geometric-operators}) as the forward model. A geodesic loss combined with multiscale locality-sensitive image similarity losses shapes the optimisation space to be smooth and largely convex in the vicinity of the solution, facilitating convergence. In later work, the same authors reduced the inference time with a rapid patient-specific network whose better initialisation needs fewer optimisation iterations to converge \cite{gopalakrishnanRapidPatientspecificNeural2025}. 
 
 \citet{downsImprovingAutomaticCerebral2025} apply the same two-stage strategy to cerebral CTA-to-DSA registration for endovascular stroke treatment, where the two modalities share few vascular structures. They register vessel segmentations rather than intensities and anchor the alignment on the large veins visible in both: a CNN trained on DRRs of the segmented CTA veins initialises the pose, which is optimised against the segmented DSA with NCC and then refined on the DSA minimum-intensity projection with MI. The pipeline registers 70\% of the 20 test patients successfully.

The network can also supply the objective rather than the starting point. \citet{gaoFullyDifferentiableFramework2024} and \citet{wangPoseawareDeepPerceptual2026} train a network that scores the similarity between the X-ray and a projection of the CT at the current pose; at test time its weights are fixed, and the pose is optimised on this learned similarity by gradient descent. Gao et al. build the network around the ProST operator (Section~\ref{sec:architectures:differentiable-geometric-operators}) and refine the result with the gradient-free CMA-ES, while Wang et al., for knee fluoroscopy, train it with a triplet loss whose margin grows with the $\mathrm{SE}(3)$ distance between poses (Section~\ref{sec:problem-formulation:classical}). In laparoscopic liver surgery, \citet{caoNonrigidImagevolumeRegistration2025} initialise the camera pose with an LSTM and then update pose and deformation together by gradient descent on a contour- and landmark-based objective rather than an image similarity.
 
\subsection{Emerging Paradigms}
\label{sec:emerging}
Implicit Neural Representations (INRs) and diffusion models also reach the transformation through many steps at inference, but neither fits the two families above: an INR is optimised from scratch for every new case, and a diffusion model follows a fixed denoising schedule. Only three of the included studies use them as the registration mechanism, two with an INR \cite{qianHybridPopulationbasedPatientspecific2025, yeNonorthogonalKVImaging2025} and one with a diffusion model \cite{kobayashiShapeRegistrationLaparoscopic2025}. This section discusses them as inference procedures; the models themselves are described in Sections~\ref{sec:arch:diffusion} and~\ref{sec:arch-inr}.

An INR represents the transformation as a continuous function of the spatial coordinates: a small MLP with weights $\psi$ takes a point $x$ as input and returns the displacement at that point, so that $\psi$ encodes the transformation $\varphi_\psi$ of a single case (Section~\ref{sec:arch-inr}). It registers a new case entirely by gradient-based optimisation at test time, applying the update of Eq.~\eqref{eq:tto}, with the same objective $J$, to the weights rather than to the transformation:
\begin{equation}
\label{eq:inr-update}
    \psi^{(j+1)} = \psi^{(j)} + U\bigl(\nabla_\psi J(\varphi_{\psi^{(j)}})\bigr).
\end{equation}
Starting from generic initial weights $\psi^{(0)}$, or from weights fitted to a network prediction as described below, the weights are updated over hundreds to thousands of steps to solve Eq.~\eqref{eq:inr}, whose solution $\psi^*$ gives the estimate $\hat\varphi = \varphi_{\psi^*}$. The main practical cost is speed. \citet{yeNonorthogonalKVImaging2025} report approximately 120 seconds per case for rigid registration, and \citet{qianHybridPopulationbasedPatientspecific2025} approximately 197 seconds for their hybrid framework. Without a population prior, the optimisation is also prone to local minima. \citet{qianHybridPopulationbasedPatientspecific2025} address both issues in two stages: a population-trained U-Net (2D3D-RegNet) first predicts a coarse DVF in a single pass, and the INR is initialised by fitting this field and then refined in a self-supervised fashion, by comparing forward projections of the deformed volume with the measured cone-beam projections. The initialisation reduces the optimisation time from about 13\,min, when the INR is fitted from scratch, to the 197\,s reported above, and lowers the risk of converging to a poor local optimum.

A diffusion model also applies a network repeatedly at inference, starting from noise and removing it over a number of denoising steps (Figure~\ref{fig:diffusion-inference}). \citet{kobayashiShapeRegistrationLaparoscopic2025}, the only included study that registers with diffusion, generate the update $\Delta\varphi$ of the registration parameters in this way, conditioned on $c = (L_s, L_p)$, the label map rendered at the initial parameters $\varphi_s$ and the labels segmented in the laparoscopic image. In the formulation of denoising diffusion probabilistic models (DDPM) \cite{ho2020denoising}, the denoising steps ${\tau = T, \dots, 1}$ compute
\begin{equation}
\label{eq:diffusion-step}
\begin{aligned}
    \Delta\varphi^{(\tau-1)} &= \mu_\theta\bigl(\Delta\varphi^{(\tau)}\bigr) + \sigma_\tau\,\eta_\tau, \\
    \mu_\theta(\Delta\varphi) &= \frac{1}{\sqrt{\alpha_\tau}}\Bigl(\Delta\varphi - \frac{1-\alpha_\tau}{\sqrt{1-\bar\alpha_\tau}}\,\epsilon_\theta(\Delta\varphi, c, \tau)\Bigr),
\end{aligned}
\end{equation}
where $\Delta\varphi^{(T)}$ and $\eta_\tau$ are drawn from a standard normal distribution, $\epsilon_\theta$ is the network that predicts the noise contained in its first argument, $\alpha_\tau$ is a fixed noise schedule with $\bar\alpha_\tau = \prod_{k=1}^{\tau}\alpha_k$, and $\sigma_\tau$ scales the sampling noise. The registered parameters are $\hat\varphi = \varphi_s + \Delta\hat\varphi$, with $\Delta\hat\varphi = \Delta\varphi^{(0)}$. Unlike an update of Eq.~\eqref{eq:multipass}, each denoising step reverses a known amount of noise rather than correcting the misalignment observed at that step, and $c$ stays the same throughout, so the label map is never re-rendered at the current estimate and the network receives no feedback on it. The other included study that uses diffusion, \citet{liDiffusionstimulatedCTUSRegistration}, applies it to domain adaptation (Section~\ref{sec:domain-adaptation}) and registers by direct regression.

\section{Neural Network Architectures}
\label{sec:architectures}
Compared to same-dimensionality registration networks, 2D-to-3D registration must explicitly address the dimensionality mismatch. Three principal strategies can be adopted to handle this challenge. The first approach performs dimensionality reduction on the 3D input, allowing for the use of well-established 2D networks. This also allows for a direct comparison of the two inputs leveraging the extensive literature in 2D image similarity. In X-ray to CT registration, for example, this can be the generation of DRRs \cite{zhengPairwiseDomainAdaptation2018, zheng2022unsupervised, daiVolumetricTumorTracking2024, quTransformerbased2D3D2024b, wangContrastiveLearningBased2024, gopalakrishnanIntraoperative2D3D2024, gaoFullyDifferentiableFramework2024, hindleyOpenSourceDeep2025, lecomte-denisDomainAgnostic2D3D2025, qianHybridPopulationbasedPatientspecific2025, yeNonorthogonalKVImaging2025, gopalakrishnanRapidPatientspecificNeural2025, yeRobustMethodReal2025, chiTwostepDeepLearning2023, zhangUnsupervised2D3DDeformable2021, miaoCNNRegressionApproach2016, downsImprovingAutomaticCerebral2025}.
Similarly, for generic SVR one can slice the 3D volume given a candidate pose \cite{dupuy2D3DDeep2023a, soormallyDeepSensorlessTracking2025, mezheritskyPopulationbased3DRespiratory2022}, but this calls for an iterative rendering during inference or for careful pose sampling during training.
The second approach employs an asymmetric or dual branch encoding, allowing for a separate processing of the 2D and 3D input that will be later fused in the latent space. This allows the original dimensionality of both inputs to be preserved while providing additional spatial context \cite{khawaledSelfattentionModelRobust2025, guoUltrasoundFrametoVolumeRegistration2023a, wuDVAPRegDualviewAnatomical2026, zhangUltrasoundSlicetoVolumeRegistration2026}. A variant of the dual-branch approach sees the 2D input within a zero-padded 3D volume enabling the processing through a unified 3D architecture \cite{chiTwostepDeepLearning2023}, or replicates the 2D feature map along depth before concatenating it with the 3D one \cite{zhangUltrasoundSlicetoVolumeRegistration2026}.

The third, and increasingly popular solution, can be interpreted as an extension of the first, integrating differentiable geometric operators directly within the network architecture: a differentiable renderer for projective imaging, or a differentiable slicing operator for SVR. By making the 2D-to-3D projection or slice extraction fully differentiable, this approach enables end-to-end learning (Section~\ref{sec:architectures:differentiable-geometric-operators}).

Beyond these fundamental design options, this field has witnessed a broad adoption of backbone architectures, from basic fully connected regressors and CNNs, towards more sophisticated transformer-based models, graph neural networks and INRs. The remainder of this section surveys the principal components employed in 2D-to-3D registration networks. We first discuss feature extraction backbones (Section~\ref{sec:architectures:backbones}), covering convolutional and transformer-based architectures that process 2D and 3D inputs into shared or comparable representations. We then examine differentiable geometric operators (Section~\ref{sec:architectures:differentiable-geometric-operators}) that encode the image geometry according to the acquisition physics within the network, enabling gradient-based optimization through the projection or slicing process. Finally, we discuss emerging generative and implicit representations (Section~\ref{sec:architectures:generative}) that offer alternative paradigms for modelling the registration problem.

\subsection{Backbones}
\label{sec:architectures:backbones}
\subsubsection{Convolutional Neural Networks}
CNNs form the foundational architecture for deep learning-based 2D-to-3D registration, first introduced to this domain by \citet{miaoCNNRegressionApproach2016}. Although more sophisticated networks have emerged, CNNs remain  widely used.

\paragraph{Encoder-Decoder architectures}
For deformable 2D-to-3D registration, where the output is a dense DVF rather than global transformation parameters, U-Net-style encoder-decoder architectures are frequently employed \cite{ronneberger2015unetconvolutionalnetworksbiomedical}. \citet{zhangUnsupervised2D3DDeformable2021} registers a 3D prior CT to a bundle of limited-angle cone-beam projections. The projections are first back-propagated into a 3D volume via a Feldkamp-Davis-Kress (FDK) reconstruction layer, and concatenated to the CT prior to a U-Net that outputs a dense DVF. The 2D-to-3D nature of the method is enforced at the loss: the deformed source is forward-projected via a differentiable Siddon layer and compared in projection space against the input cone-beam projections. \citet{tianLiftRegLimitedAngle2022} also back-project limited-angle projections into 3D, but replace the decoder with fully connected layers that regress the coefficients of a PCA model of lung deformations.
\citet{mezheritskyPopulationbased3DRespiratory2022} combine a pretrained U-Net and a dual branch encoder to first have a properly pretrained DVF decoder and then fine tune on additional 2D information.

\begin{figure}[!t]
    \centering
    \includegraphics[width=0.75\textwidth]{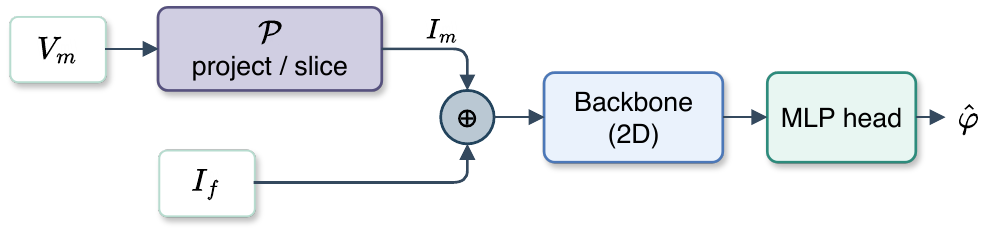}
    \caption{Single-stream encoder. $V_m$ is first reduced to a 2D image $I_m$ by a spatial transformer, then concatenated (``+'') with $I_f$ and processed by one backbone; the task becomes 2D-to-2D registration at the cost of volumetric context.}
    \label{fig:single-stream}
\end{figure}

\paragraph{Single-Stream Encoders}
Standard image classification backbones, particularly ResNet variants \cite{he2015deepresiduallearningimage}, are widely used for simple 2D and 3D feature extraction. A common design specific for projective imaging is DRR and X-ray-based registration. This approach typically uses concatenated inputs as shown in Figure \ref{fig:single-stream} \cite{miaoCNNRegressionApproach2016, zhengPairwiseDomainAdaptation2018, zhangUnsupervised2D3DDeformable2021, zheng2022unsupervised, chiTwostepDeepLearning2023, gopalakrishnanIntraoperative2D3D2024, gaoFullyDifferentiableFramework2024, hindleyOpenSourceDeep2025, lecomte-denisDomainAgnostic2D3D2025, gopalakrishnanRapidPatientspecificNeural2025}.  

\begin{figure}[!t]
    \centering
    \includegraphics[width=0.6\textwidth]{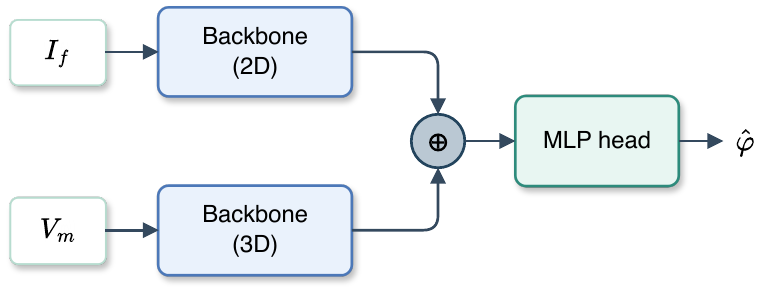}
    \caption{Dual-branch encoder. $I_f$ and $V_m$ are embedded by separate backbones and fused (``+'') before the MLP, so each input keeps its native dimensionality until fusion. \citet{yeRobustMethodReal2025} use this design with a projection of $V_m$.}
    \label{fig:dual-branch}
\end{figure}

\paragraph{Dual Branch Configurations}
Dual-branch neural networks characterized by two independent branches that embed the input (see Figure \ref{fig:dual-branch}). For the 2D-to-3D registration case these are $I_f$ and $V_m$, or $I_f$ and a slice or projection of $V_m$. Branches can also separate input representations rather than images: \citet{zhangEnhancingCNNRegressors2026} process the intensity and binary-contour versions of the X-ray--DRR pair in two streams fused by channel attention. Although Dual-branch networks do not necessarily require the branches to process inputs of different dimensionality, it is the most common way to allow such input, including a pairing of per-view CNNs with a graph network on the vertebra's surface mesh \cite{wuDVAPRegDualviewAnatomical2026} (Section~\ref{sec:direct}). \citet{wu3DIsotropicHighResolution2025} designed a dual branch network for SVR of MRI slices to their own MRI volume by computing input embeddings of the MRI slice of interest and the MRI volume on distinct embedding branches, to be later fused within the latent space. \citet{hindleyOpenSourceDeep2025} investigated the best moment to fuse inputs. They compared an early fusion (concatenation of X-ray and DRR), a dual-branch of 2D inputs (X-ray and DRR), and a dual-branch of a 2D and 3D inputs (X-ray and CT). Early fusion demonstrated the lowest centroid error out of the three. 

\subsubsection{Vision Transformers and Attention-based Mechanisms}
\label{sec:architectures:attention}
Transformers and, more generally, attention-based mechanisms have emerged in popularity in  recent years. These methods bring a different set of inductive biases compared to convolutional architectures, and they possess desirable properties for 2D-to-3D registration. For $M$ input tokens (image patches or slices) stacked in $X \in \R^{M \times d_x}$, self-attention computes
\begin{equation}
\label{eq:attention}
    \mathrm{Attn}(X) = \operatorname{softmax}\biggl(\frac{(XW_Q)(XW_K)^{\top}}{\sqrt{d_k}}\biggr)\, XW_V,
\end{equation}
with learned projections $W_Q$, $W_K$, $W_V$ and key dimension $d_k$. Every output token is thus a weighted combination of all input tokens, through an $M \times M$ attention matrix. The advantages of these biases are twofold: their global context captures long-range spatial dependencies that CNNs miss due to limited receptive fields, while the sequence-processing nature of transformers naturally accommodates multiple slices or views as a sequence, which is a structural advantage for multi-slice SVR and multi-view registration \cite{xuSVoRTIterativeTransformer2022, khawaledSelfattentionModelRobust2025}.

However, transformers lack strong visual inductive biases which is inherent to CNNs, resulting in lower data efficiency, hence requiring larger amounts of data to train, often requiring pretraining \cite{dosovitskiy2021imageworth16x16words}. 
The global attention mechanism of transformers is also generally computationally demanding, with $O(M^2)$ complexity over the input size, often infeasible for high-resolution images, compared to faster $O(M)$ CNN-based approaches \cite{vaswani2023attentionneed, dosovitskiy2021imageworth16x16words}. Efforts have been made to reduce such complexity by avoiding global attention, for example with Swin Transformers \cite{liu2021swintransformerhierarchicalvision} which, several studies included in this review have opted for \cite{daiVolumetricTumorTracking2024, quTransformerbased2D3D2024b, wangContrastiveLearningBased2024, yeRobustMethodReal2025}.
Pure transformer architectures remain rare; most adopt hybrid transformer-CNN designs where CNNs handle low level feature extraction and transformers model high-level feature relationships.

\paragraph{Transformers as Feature Extraction Backbones} 
Transformers can act as substitutes of CNN backbones when global context is beneficial. Nevertheless, efforts have been made to exploit their inherit advantages. \citet{yeRobustMethodReal2025} proposed a rigid registration network for CT and X-rays using a Swin Transformer combined with Squeeze-and-Excitation (SE) Attention modules \cite{hu2019squeezeandexcitationnetworks} and Feature Pyramid Networks \cite{lin2017featurepyramidnetworksobject}, achieving sub-millimeter alignments over 3 different anatomies. 

\paragraph{Transformer for Multi-Element Slice Sequence Processing}
Transformers can be exploited for their sequence processing nature. In the 2D-to-3D registration setting, stacks of 2D slices or frames can be naturally interpreted as ordered sequences, allowing attention mechanisms to capture inter-slice dependencies that are difficult to encode with purely convolutional architectures. This is particularly useful for non-projective data, where a single 2D sample does not hold information about the z-axis.
\citet{xuSVoRTIterativeTransformer2022} use self-attention to share information across slices to predict rigid transformations. They introduced a positional embedding that encodes both the stack index (which acquisition stack does a slice belong to) and the slice index (its position within a stack), encoding spatial relationships. 
\citet{khawaledSelfattentionModelRobust2025} implement a self-attention transformer that predicts pixel-wise score maps for each MRI slice before dual-branch encoding, effectively decreasing sensitivity to noise.

\paragraph{Transformers with Geometric Representations} 
Recent work combines transformer modules with complementary representation types to capture different aspects of the registration problem. \citet{quTransformerbased2D3D2024b} propose GDSwin-t, a dual-branch architecture for X-Ray and CT spine registration that combines Swin Transformers with Graph Neural Networks (GNNs). The first branch extracts anatomical feature points using a ConvNeXt network \cite{liu2022convnet2020s} and constructs a graph over these landmarks, processing it through GraphSAGE \cite{hamilton2018inductiverepresentationlearninglarge} aggregation and DiffPool \cite{ying2019hierarchicalgraphrepresentationlearning} hierarchical pooling to produce an anatomical embedding that captures the structural regularity of the vertebral column. The second branch employs a four-stage Swin Transformer with deformable attention to extract multi-scale global features \cite{xia2022visiontransformerdeformableattention}. The two representations are fused to regress the affine transformation matrix.

\subsection{Differentiable Geometric Operators}
\label{sec:architectures:differentiable-geometric-operators}
Both forms of the operator $\mathcal P$ in Eq.~\eqref{eq:reg_crossdim} can be written as an evaluation of the warped volume $V = \mathcal W(V_m, \varphi)$ at each pixel $u$ of the 2D image:
\begin{equation}
\label{eq:projection-operator}
\mathcal P(V)(u) =
\begin{cases}
\displaystyle\int_{r_u} V(x)\,\mathrm dx, \\[8pt]
V(u_1, u_2, 0),
\end{cases}
\end{equation}
for projective and planar imaging, respectively. The projective form is the line integral of the attenuation coefficients along the ray $r_u$ from the X-ray source to the position of pixel $u$ on the detector; the planar form samples the volume on the plane $z=0$. In both cases, $\varphi$ enters only through the positions at which $V_m$ is evaluated, so the operator is differentiable with respect to the pose whenever this sampling is. The operators below differ in how they discretise the integral and interpolate the volume.

\subsubsection{Differentiable Rendering and Projection}
In projection-based 2D-to-3D registration, forward models have been traditionally non-differentiable with respect to pose parameters \cite{goitein1983multi, markelj2012review}. This prevented end-to-end learning through the projection process, forcing early methods to treat DRR generation as a fixed preprocessing step rather than an integrated network component; physically realistic simulators such as DeepDRR \cite{unberath2018deepdrr} were likewise designed to generate training data rather than to propagate gradients to the pose. Recent advancements proposed fast and differentiable solutions \cite{gaoFullyDifferentiableFramework2024, gopalakrishnan2022fast} enabling self-supervised learning by comparing rendered DRRs to target images without ground-truth labels, and gradient-based test-time optimization that converges faster and more reliably than gradient-free alternatives such as CMA-ES \cite{hansen2023cmaevolutionstrategytutorial}.

\citet{gaoFullyDifferentiableFramework2024} introduced the Projective Spatial Transformer (ProST), extending the spatial transformer framework of \citet{jaderberg2016spatialtransformernetworks} to projective X-ray transmission imaging. ProST defines a sampling grid of control points along rays emanating from a virtual X-ray source following a cone-beam geometry. Given a 6-DoF rigid pose $\varphi_r$ in the special Euclidean group $\mathrm{SE}(3)$ (Section~\ref{sec:problem-formulation:classical}), the grid is rigidly transformed. Trilinear interpolation samples CT voxel values at the transformed control point locations, and a ray-summation operation collapses the sampled values to produce a DRR. All steps are differentiable, enabling  gradients to flow from a learned 2D similarity score back through the projection to the 3D pose parameters.

The DiffDRR renderer of \citet{gopalakrishnan2022fast} takes a different approach from ProST. Rather than sampling the volume at control points along each ray, it traces each ray with Siddon's algorithm \cite{siddon1985fast}, which weights every voxel by the exact length of the ray crossing it. Because this is implemented as vectorised tensor operations, the rendered X-ray is differentiable with respect to the rigid pose. \citet{gopalakrishnanIntraoperative2D3D2024} use DiffDRR both to generate unlimited patient-specific training images and as the forward model of the test-time optimisation in DiffPose (Section~\ref{sec:hybrid-tto}).

\citet{jiangSemanticallyConsistentDeformable2D3D2025a} extend differentiable rendering to deformable 2D-to-3D registration for craniofacial structure estimation from a single lateral cephalometric radiograph. Their coarse-to-fine framework first regresses low-dimensional coordinates in a statistical registration field subspace (built via PCA of deformation fields from clinical CBCTs), then refines the resulting coarse field at the voxel level using 2D feature embeddings projected back into 3D via the known projection geometry. The differentiable volume renderer generates synthetic radiographs of the deformed template volume, enabling a self-supervised image similarity loss that eliminates the need for paired volumetric training data.

\subsubsection{Differentiable Spatial Transformations}
While differentiable rendering addresses the projection-based modalities, slice-based modalities  (US, MRI) registration requires an analogous capability of differentiability extracting a 2D plane from a 3D volume at a predicted pose. The spatial transformer network (STN) framework from \citet{jaderberg2016spatialtransformernetworks} provides this foundation: given predicted transformation parameters, a sampling grid is generated and trilinear interpolation extracts the corresponding 2D slice from the volume. This operation is differentiable with respect to both the transformation parameters and the volume contents. An early use of this layer for learning-based registration, although between 2D images, is DIRNet \cite{devos2017endtoend}, which couples a convolutional regressor with a spatial transformer and a resampler and is trained end to end on an image similarity loss alone.

\citet{guoUltrasoundFrametoVolumeRegistration2023a} proposes a straightforward self-supervised approach using a spatial transformer to extract predicted slices from a reconstructed US volume, comparing them with observed US frames to verify and correct frame-to-volume registration estimates. \citet{khawaledSelfattentionModelRobust2025} similarly employ a 3D slice grid generator based on predicted rigid parameters to reconstruct the slices and compute the grid-based loss during training, rather than relying on intensity-level comparison.

\subsection{Generative and Implicit Representations}
\label{sec:architectures:generative}
\subsubsection{Diffusion Models}
\label{sec:arch:diffusion}
\begin{figure}[!t]
    \centering
    \includegraphics[width=0.7\textwidth]{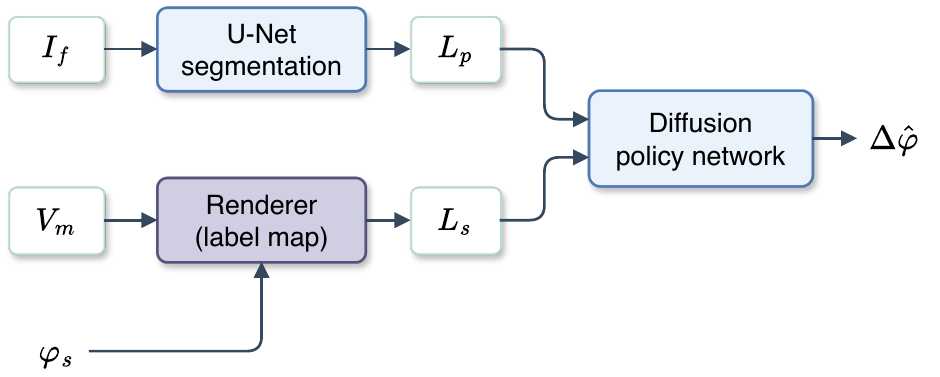}
    \caption{Inference with the diffusion policy of \citet{kobayashiShapeRegistrationLaparoscopic2025}. A U-Net segments the laparoscopic frame $I_f$ into a label map $L_p$, and a renderer produces the label map $L_s$ of the liver mesh $V_m$ at the initial parameters $\varphi_s$. Conditioned on both label maps, the policy network predicts the parameter update $\Delta\hat\varphi$ through its denoising steps; the registered parameters are $\varphi_s + \Delta\hat\varphi$.}
    \label{fig:diffusion-inference}
\end{figure}

Since 2020, Diffusion models have emerged as powerful generative frameworks in computer vision \cite{ho2020denoising}. However, only later they were applied to same-dimensionality registration exploring their use for producing continuous deformation fields or conditional score functions \cite{kazerouni2023diffusion}.
Their adoption in 2D-to-3D registration remains scarce, with only two works included in this review. \citet{kobayashiShapeRegistrationLaparoscopic2025} proposed the first application of diffusion learning to 2D-to-3D shape registration, using an offline diffusion policy network to register preoperative liver surface meshes to intraoperative laparoscopic images. Instead of adding noise to the image space, their method adds Gaussian noise to seven registration parameters (the orientation of the mesh, its position in the image plane and a uniform scale) and trains the network to remove it. The network never sees the laparoscopic image directly: its input is a nine-channel image made of the semantic labels rendered at two consecutive noise levels and a positional encoding, so that a model trained on rendered labels can be applied to labels segmented from real images. The inference pipeline is shown in Figure~\ref{fig:diffusion-inference} and discussed in Section~\ref{sec:emerging}.

\citet{liDiffusionstimulatedCTUSRegistration} use a diffusion model in a different role, for domain adaptation (Section~\ref{sec:domain-adaptation}): it converts real ultrasound images into the style of ultrasound simulated from CT, so that a regression network trained on simulated images can be applied to real ones.

\subsubsection{Implicit Representation Networks}
\label{sec:arch-inr}
INRs represent a fundamentally different paradigm from the data-driven architectures discussed above. Rather than learning a mapping from input images to output predictions across a training dataset, an INR encodes a single signal (e.g. an image, volume, DVF) entirely within the weights of a compact MLP, and are optimized per instance without requiring external datasets. For registration, the signal is the transformation itself: an MLP $f_\psi$ with weights $\psi$ maps each coordinate $x\in\R^3$ to a displacement, $\varphi_\psi(x) = x + f_\psi(x)$, and the weights are fitted to the case at hand by maximising the objective $J$ of Eq.~\eqref{eq:tto}, the registration objective of Eq.~\eqref{eq:reg_crossdim}, over $\psi$:
\begin{equation}
\label{eq:inr}
    \psi^* = \argmax_{\psi}\; J(\varphi_\psi).
\end{equation}
Compared with test-time optimisation (Section~\ref{sec:hybrid-tto}), the objective is the same, but the variable is the weights of a network rather than the transformation parameters, and no network is trained beforehand. The fitted MLP is a continuous, differentiable representation of the field that can be queried at any coordinate. Since each INR is fitted to a single case, there is no training set, and hence no distribution shift between training and test data and no sensitivity to variability in scan configurations or patient anatomy across a population, at the cost of slow optimisation and a higher risk of local minima (Section~\ref{sec:emerging}).

The properties of the fitted field depend largely on the activation function. MLPs with ReLU activations are biased towards low-frequency functions \cite{rahaman2019spectralbiasneuralnetworks}, which favours smooth fields but limits fine detail; Fourier features of the input coordinates \cite{tancik2020fourier} or sine activations, as in SIREN \cite{sitzmann2020siren}, remove this bias, and a scale or frequency factor $\omega_0$ sets the highest frequency the field can represent. The activation also determines which regularisers are available. A ReLU network is piecewise linear, so its second derivatives vanish and a bending-energy penalty is zero everywhere, whereas the derivatives of a sine network are themselves sine networks, so first- and second-order penalties, including the Jacobian-determinant terms that discourage folding, can be computed exactly by automatic differentiation \cite{wolterink2022implicit, vanharten2024cycle}. These penalties remain necessary: an unregularised SIREN produced several large errors on lung CT \cite{wolterink2022implicit}, and large values of $\omega_0$ were associated with folding of the field \cite{byra2023inr}. Diffeomorphism can instead be built into the representation by letting the MLP encode a stationary velocity field that is integrated by scaling and squaring \cite{han2023dnvf}.

\begin{figure*}[!t]
    \centering
    \includegraphics[width=0.8\textwidth]{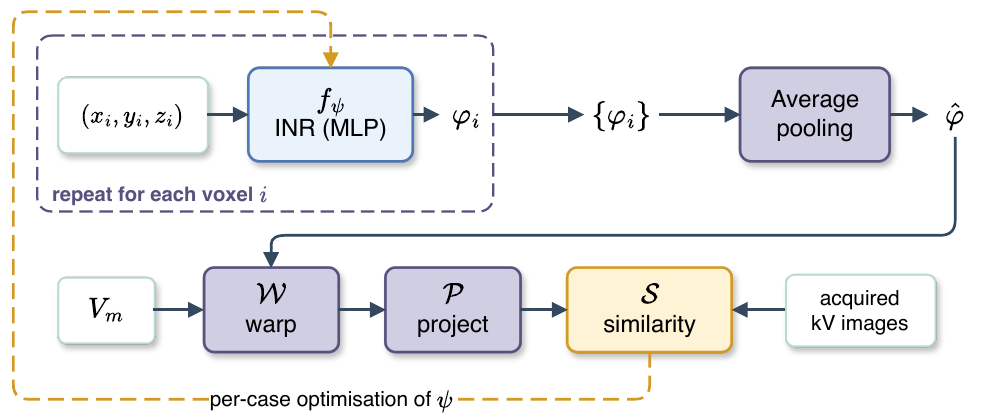}
    \caption{RegiNet \citep{yeNonorthogonalKVImaging2025}. An MLP maps each voxel coordinate of $V_m$ to a rigid estimate $\varphi_i$; average pooling over all voxels yields the global estimate $\hat\varphi$. The MLP is optimised per case, without a training set (Section~\ref{sec:arch-inr}).}
    \label{fig:inr}
\end{figure*}

\citet{yeNonorthogonalKVImaging2025} proposed a dataset-free INR framework for 2D-to-3D rigid registration in non-coplanar radiation therapy (shown in Figure \ref{fig:inr}). Their RegiNet consists of a three-layer MLP that maps voxel coordinates to per-voxel motion vectors, followed by average pooling to enforce rigid-body consistency and recover a single 6-DoF motion parameter set.

\citet{qianHybridPopulationbasedPatientspecific2025} employ INRs in a complementary fashion for deformable 2D-to-3D registration in limited-angle CBCT estimation. Their 2D3D-INR is a five-layer MLP with sine activations that takes voxel coordinates as input, outputs the DVF at those coordinates and is fitted with an explicit smoothness term; within the HB-2D3DReg framework, it refines the coarse DVF of a population-trained U-Net (Section~\ref{sec:emerging}).

\section{Quantitative Metrics}
\label{sec:eval}
Evaluating 2D-to-3D registration performance presents several challenges. The absence of proper ground truth allows systematic bias to propagate from the algorithms used to generate such labels. Secondly, given the dimensional mismatch between the data being compared, there is no single natural space in which to measure error.
A third complication is that many of the reported quantities at evaluation can also serve as training objectives. This dual role blurs the line between optimisation target and independent assessment, and its implications are revisited in the relevant sections below. The metrics used across the reviewed studies can be grouped into five categories:
\begin{description}
    \item[Geometric accuracy metrics] assess spatial alignment of the registered $V_m$ in either 3D space or the 2D image plane (requiring projection or slicing).
    \item[Transformation parameter metrics] quantify errors directly in the transformation domain, either on rigid pose parameters or voxel-wise on displacement fields for deformable registration.

    \item[Overlap and shape metrics] measure the area correspondence between the aligned anatomical structures.
    \item[Image similarity metrics] evaluate alignment in the intensity domain, providing a complementary perspective, which is particularly useful for self-supervised methods.
    \item[Robustness and efficiency metrics] assess reliability and computational performance.
\end{description}
Table~\ref{tab:metrics-summary} lists, for each included study, the metrics it reports under these categories.
\begin{table*}[!htbp]
\centering
\caption{Evaluation metrics reported by the reviewed articles.}
\label{tab:metrics-summary}
\footnotesize
\newcommand{\rot}[1]{\rotatebox{90}{\textbf{#1}}}
\setlength{\tabcolsep}{9pt}
\begin{tabular}{l cccccccccc l}
\toprule
\textbf{Article} & 
\rot{mTRE (3D)} & 
\rot{mTRE (2D)} & 
\rot{Parameter error} & 
\rot{DVF error} & 
\rot{DSC} & 
\rot{Surface distance} & 
\rot{Image similarity} & 
\rot{Capture range} & 
\rot{Success rate} & 
\rot{Time} & 
\textbf{Other} \\
\midrule
    \citet{caoNonrigidImagevolumeRegistration2025} &  &  & \checkmark &  &  & \checkmark &  &  &  & \checkmark & -- \\
    \citet{chiTwostepDeepLearning2023} &  &  &  &  &  & \checkmark &  & \checkmark &  &  & -- \\
    \citet{daiVolumetricTumorTracking2024} &  & \checkmark &  &  & \checkmark &  &  &  &  & \checkmark & CE \\
    \citet{downsImprovingAutomaticCerebral2025} &  & \checkmark &  &  &  &  &  & \checkmark & \checkmark &  & -- \\
    \citet{dupuy2D3DDeep2023a} & \checkmark &  &  &  &  &  & \checkmark &  &  & \checkmark & drift \\
    \citet{facenteRealtimeAlignment3D2026} & \checkmark & \checkmark & \checkmark &   &   &   &   &   &   & \checkmark & -- \\
    \citet{gaoFullyDifferentiableFramework2024} & \checkmark &  & \checkmark &  &  &  &  & \checkmark & \checkmark & \checkmark & -- \\
    \citet{gopalakrishnanIntraoperative2D3D2024} &  & \checkmark &  &  &  &  &  &  & \checkmark & \checkmark & -- \\
    \citet{gopalakrishnanRapidPatientspecificNeural2025} & \checkmark & \checkmark & \checkmark &  &  &  &  &  & \checkmark & \checkmark & AUC \\
    \citet{guoUltrasoundFrametoVolumeRegistration2023a} & \checkmark &  &  &  &  &  & \checkmark &  &  & \checkmark & drift \\
    \citet{hindleyOpenSourceDeep2025} &  &  &  &  & \checkmark &  & \checkmark &  &  & \checkmark & CE, Jac. \\
    \citet{jiangSemanticallyConsistentDeformable2D3D2025a} &  &  &  &  & \checkmark &  & \checkmark &  &  & \checkmark & CM \\
    \citet{khamenehAutomaticCNNbased3D2025} & \checkmark &  &  &  &  & \checkmark &  &  & \checkmark & \checkmark & CM \\
    \citet{khawaledSelfattentionModelRobust2025} & \checkmark &  & \checkmark &  &  &  &  &  &  & \checkmark & -- \\
    \citet{kobayashiShapeRegistrationLaparoscopic2025} &  &  & \checkmark &  &  &  &  &  &  &  & -- \\
    \citet{lecomte-denisDomainAgnostic2D3D2025} &  &  &  &  & \checkmark &  &  &  &  & \checkmark & -- \\
    \citet{liDiffusionstimulatedCTUSRegistration} &  &  & \checkmark &  &  &  & \checkmark &  &  & \checkmark & -- \\
    \citet{mezheritskyPopulationbased3DRespiratory2022} & \checkmark &  &  &  &  &  & \checkmark &  &  & \checkmark & Jac. \\
    \citet{mhiriNeuralPatientspecific3D2D2025b} & \checkmark & \checkmark &  &  &  &  &  &  &  & \checkmark & vRMSE \\
    \citet{miaoCNNRegressionApproach2016} &  & \checkmark & \checkmark &  &  &  &  & \checkmark & \checkmark & \checkmark & prec. \\
    \citet{mohsenisalehiRealTimeDeepPose2019} &  &  & \checkmark &  &  &  &  & \checkmark &  & \checkmark & -- \\
    \citet{oya2D3DDeformable2024} & \checkmark &  &  &  & \checkmark & \checkmark &  &  &  & \checkmark & -- \\
    \citet{qianHybridPopulationbasedPatientspecific2025} & \checkmark &  &  &  &  &  & \checkmark &  &  & \checkmark & -- \\
    \citet{quTransformerbased2D3D2024b} & \checkmark &  & \checkmark &  &  &  & \checkmark &  & \checkmark & \checkmark & -- \\
    \citet{soormallyDeepSensorlessTracking2025} & \checkmark &  & \checkmark &  &  &  &  &  & \checkmark & \checkmark & drift \\
    \citet{tianLiftRegLimitedAngle2022} & \checkmark &  &  &  & \checkmark &  &  &  &  & \checkmark & -- \\
    \citet{wangContrastiveLearningBased2024} &  & \checkmark & \checkmark &  &  &  & \checkmark &  &  & \checkmark & -- \\
    \citet{wangPoseawareDeepPerceptual2026} &   &   & \checkmark &   &   &   &   & \checkmark & \checkmark &   & -- \\
    \citet{weiDeepLearningApproach2021} &  &  & \checkmark &  &  & \checkmark &  & \checkmark &  & \checkmark & -- \\
    \citet{wu3DIsotropicHighResolution2025} &  &  &  &  &  &  &  &  &  & \checkmark & recon. \\
    \citet{wuDVAPRegDualviewAnatomical2026} &   &   & \checkmark &   &   & \checkmark &   &   &   & \checkmark & -- \\
    \citet{xuSVoRTIterativeTransformer2022} & \checkmark &  & \checkmark &  &  &  & \checkmark &  &  & \checkmark & recon. \\
    \citet{yeNonorthogonalKVImaging2025} &  &  & \checkmark &  &  &  &  &  &  & \checkmark & -- \\
    \citet{yeRobustMethodReal2025} & \checkmark &  & \checkmark &  &  &  &  &  & \checkmark & \checkmark & -- \\
    \citet{zhangUnsupervised2D3DDeformable2021} & \checkmark &  &  &  &  &  & \checkmark &  &  & \checkmark & -- \\
    \citet{zhangDeformableRegistrationLateral2021} &  & \checkmark &  & \checkmark &  & \checkmark & \checkmark &  &  & \checkmark & -- \\
    \citet{zhangEnhancingCNNRegressors2026} & \checkmark &   & \checkmark &   & \checkmark &   &   &   &   & \checkmark & -- \\
    \citet{zhangUltrasoundSlicetoVolumeRegistration2026} & \checkmark &   & \checkmark &   & \checkmark &   & \checkmark & \checkmark &   & \checkmark & -- \\
    \citet{zhangWarmstartedReinforcementLearning2026} & \checkmark & \checkmark &   &   &   &   &   &   &   & \checkmark & -- \\
    \citet{zhengPairwiseDomainAdaptation2018} & \checkmark &  & \checkmark &  &  &  &  &  & \checkmark &  & -- \\
    \citet{zheng2022unsupervised} & \checkmark &  &  &  &  &  &  &  & \checkmark & \checkmark & -- \\
\midrule
\textbf{Studies (of 41)} & \textbf{22} & \textbf{10} & \textbf{21} & \textbf{1} & \textbf{8} & \textbf{7} & \textbf{13} & \textbf{8} & \textbf{12} & \textbf{36} & \\
\bottomrule
\end{tabular}
\vspace{2pt}
\begin{flushleft}
\scriptsize
\textbf{Legend:} \textbf{mTRE 3D}: mean target or landmark registration error in 3D; \textbf{mTRE 2D}: error measured in the image plane after projection; \textbf{Param.\ error}: error on rigid or affine parameters; \textbf{DVF error}: voxel- or control-point-wise error against a reference displacement field; \textbf{DSC}: Dice or IoU overlap; \textbf{Surface dist.}: Hausdorff, mean surface, contour or plane distances; \textbf{Image sim.}: NCC, MI, SSIM, PSNR or intensity error reported as an evaluation metric; \textbf{Capture range}: accuracy or success as a function of the initial misalignment; \textbf{Success rate}: fraction of cases under an error threshold, or failure rate; \textbf{Time}: inference or registration time. \textbf{Other}: CE = tumour or target centroid error; Jac. = Jacobian determinant of the deformation; CM = error of derived clinical measurements; drift = error accumulation along a sequence; prec.\ = spread of repeated registrations from perturbed initialisations; recon.\ = quality of the reconstructed volume; AUC = area under the success-rate curve over thresholds; vRMSE = vertex-wise RMSE of the deformed mesh.
\end{flushleft}
\end{table*}

\subsection{Geometric accuracy}
\label{sec:eval:geometrical}
The Mean Target Registration Error (mTRE) is the mean Euclidean distance between $N$ target points $x_i$ of $V_m$ mapped by the predicted and the ground-truth transformation,
\begin{equation}
\label{eq:mtre}
    \mathrm{mTRE} = \frac{1}{N}\sum_{i=1}^{N} \bigl\lVert \hat\varphi(x_i) - \varphi_\mathrm{GT}(x_i) \bigr\rVert_2,
\end{equation}
and it is the most widely adopted metric: 28 of the 41 included studies report it in its 3D or projected form (Table~\ref{tab:metrics-summary}), 22 of them in 3D. 
However, the placement of these landmarks brings substantial heterogeneity in how these target points are defined. In addition, because mTRE depends on the spatial distribution of the chosen targets, landmarks near the rotation centre exhibit smaller errors than peripheral points for the same pose error, complicating cross-study mTRE value comparison.

Several studies instead evaluate registration accuracy in the 2D image plane. \citet{miaoCNNRegressionApproach2016} introduced the projected target registration error, which replaces both points in Eq.~\eqref{eq:mtre} by their projections $\pi(\cdot)$ onto the detector,
\begin{equation}
\label{eq:mtre-proj}
    \mathrm{mTRE}_\mathrm{proj} = \frac{1}{N}\sum_{i=1}^{N} \bigl\lVert \pi\bigl(\hat\varphi(x_i)\bigr) - \pi\bigl(\varphi_\mathrm{GT}(x_i)\bigr) \bigr\rVert_2.
\end{equation} Other works introduced similar projection-domain metrics under different names: \citet{downsImprovingAutomaticCerebral2025} introduced Mean Projection Error (MPE), \citet{mhiriNeuralPatientspecific3D2D2025b} used Landmark Reprojection Distance (LRD), and \citet{gopalakrishnanIntraoperative2D3D2024} defined a projected mTRE using the camera intrinsic matrix for landmark projection. 

Projection-based errors are inherently insensitive to displacements along the projection axis. Under perspective projection, a target at depth $z$ from the source and lateral distance $x$ from the central ray moves in the image by the same amount under a depth error $\Delta z$ as under an in-plane error of $(x/z)\,\Delta z$, and $x/z$ is small because targets lie close to the central ray and far from the source. The evaluation of LXPose by \citet{facenteRealtimeAlignment3D2026} illustrates the consequence: the mean projection distance is below 1\,mm on both test datasets, while the 3D mTRE exceeds 11\,mm, mostly along the depth axis. Projected errors should therefore be considered complementary to 3D mTRE when volumetric spatial accuracy is relevant.

\subsection{Parameter-wise evaluation}
\label{sec:eval:parameter-wise}
When ground-truth pose parameters are available, performance is frequently quantified using parameter-wise errors such as Mean Absolute Error (MAE) or Root Mean Square Error (RMSE) over the 6-DoF rigid transformation \cite{oya2D3DDeformable2024, daiVolumetricTumorTracking2024, wangContrastiveLearningBased2024, mhiriNeuralPatientspecific3D2D2025b}. Most works further decompose the error into separate translation and rotation components, since these quantities have different units and different geometric interpretations.
This decomposition requires choosing how to represent and measure rotation errors, and the reviewed literature shows considerable diversity in this choice. The most common approach is to represent rotations via Euler angles and report MAE in degrees for each axis independently, a straightforward choice when rotations are small \cite{miaoCNNRegressionApproach2016, guoUltrasoundFrametoVolumeRegistration2023a, yeRobustMethodReal2025}. Larger rotations represented using Euler angles may incur in gimbal lock and encounter discontinuities that might hinder the flow of gradients. \citet{mohsenisalehiRealTimeDeepPose2019} proposes using the axis-angle representation, and propose measuring rotation error via the geodesic distance on $\mathrm{SO}(3)$ of Eq.~\eqref{eq:geodesic}. The choice of representation is not merely a notational convenience: it also affects how well a pose can be optimised (Section~\ref{sec:hybrid-tto}), and it deserves more attention than it has received in the literature.

In the context of deformable registration, the analogous evaluation compares the predicted DVF $\hat{\varphi}_d$ against a reference field $\varphi_d^{\mathrm{GT}}$, which is known by construction when the deformation is simulated and must otherwise be estimated, typically by a classical registration between volumes. The most common choices are the voxel-wise MSE or endpoint error (the mean Euclidean norm $\|\hat{\varphi}_d(x) -\varphi_d^{\mathrm{GT}}(x)\|_2$ averaged over the image domain). In the reviewed literature this form of evaluation is rare: only \citet{zhangDeformableRegistrationLateral2021} report an error against a reference displacement field (Table~\ref{tab:metrics-summary}), computed by B-spline registration of clinical CBCTs, as voxel-dense reference fields cannot be directly measured in clinical practice (see Section~\ref{sec:supvsunsup:sup}). The other deformable methods rely instead on landmark errors, image-similarity and overlap-based metrics (see Sections \ref{sec:eval:image-similarity}, \ref{sec:eval:overlap-shape}) as their primary evaluation channel.

Although parameter-wise errors provide a well-defined and directly interpretable measurement of registration accuracy, they are only an indirect proxy for anatomical alignment. The relationship between pose error and displacement of anatomical structures is geometry dependent, a small rotation produces displacements that scale with the distance from the rotation center. 
For these reasons, parameter-wise evaluation is best used alongside, and not in place of, target-based geometric metrics.

\subsection{Overlap and Shape Metrics}
\label{sec:eval:overlap-shape}
Overlap metrics assess registration through the correspondence of segmented anatomical structures. The Dice Similarity Coefficient (DSC), reported by eight of the 41 studies (Table~\ref{tab:metrics-summary}), is the most prevalent. It is computed in 3D on the deformed volume or in 2D on projected or sliced masks, mostly in deformable registration \cite{daiVolumetricTumorTracking2024, hindleyOpenSourceDeep2025, oya2D3DDeformable2024}, and is occasionally also used as a training objective through its differentiable soft-Dice formulation \cite{daiVolumetricTumorTracking2024, lecomte-denisDomainAgnostic2D3D2025}.

DSC values must be interpreted with care, because their sensitivity to misalignment depends on the size and shape of the structure. \citet{lecomte-denisDomainAgnostic2D3D2025} show that a translation of about 7\,mm from perfect alignment lowers the DSC of thin structures such as blood vessels to 0.65, but that of the whole liver only to 0.92, and in deformable registration \citet{rohlfing2012surrogates} found that the overlap of large structures could not separate accurate from inaccurate registrations, whereas that of small, localised structures could. Symmetry adds a further ambiguity. A structure whose mask is symmetric about one axis gives the same overlap for its true pose and for the pose rotated by 180\textdegree{} about that axis, so DSC has two equally good optima, one of them wrong; with symmetry about two axes there are four, of which only one is correct. For nearly cylindrical structures, such as a long bone or a vessel segment, a rotation about the long axis barely changes the overlap at all. When masks are compared after projection, DSC also inherits the depth insensitivity discussed in Section~\ref{sec:eval:geometrical}, since a structure displaced along the projection ray covers almost the same pixels.

Surface distances complement overlap by measuring discrepancies between boundaries rather than volumes. \citet{oya2D3DDeformable2024} report the Hausdorff Distance (HD), which captures the worst-case boundary discrepancy and is therefore sensitive to single outlying points. \citet{zhangDeformableRegistrationLateral2021} introduced Mean Contour Deviation (MCD) and Surface Mesh Deviation (SMD), comparing evenly sampled points between the generated DRR and the reference lateral cephalogram, and boundary meshes extracted automatically with ITK-SNAP \cite{itksnap}, respectively. Following metric-selection guidelines for biomedical image analysis \cite{maierhein2024metrics}, overlap is most informative for structures of a size at which misalignment changes it appreciably, and is best paired with a boundary distance.

\subsection{Image Similarity Metrics}
\label{sec:eval:image-similarity}
Thirteen of the 41 studies report image-domain similarity metrics, including Normalised Cross-Correlation (NCC), Normalised Mutual Information (NMI), Structural Similarity Index Measure (SSIM), and Peak Signal-to-Noise Ratio (PSNR), as supplementary evaluation measures (Table~\ref{tab:metrics-summary}), computed between the fixed image and the registered projection or slice, or between the registered and a reference volume. They need no annotation, but their absolute values depend on the imaging modality, resolution and preprocessing pipeline of each study, which makes inter-study comparison unreliable. Similarity values also carry limited information about the pose, particularly across modalities, where the same anatomy produces different intensities. In deformable registration the concern is sharper: \citet{rohlfing2012surrogates} constructed transformations that are highly inaccurate yet score well on image similarity and tissue overlap, because a sufficiently flexible field can match intensities through an anatomically implausible deformation. Of the six deformable studies that report image similarity, only \citet{hindleyOpenSourceDeep2025} and \citet{mezheritskyPopulationbased3DRespiratory2022} also report the Jacobian determinant of the field, which would reveal such folding.

Image similarity metrics are also widely used as training objectives, particularly in self-supervised pipelines where they constitute the sole learning signal. In most studies the exact metric reported at evaluation differs from the one minimised during training, so intensity-based evaluation still provides a channel that is at least partially independent of the training objective. A circularity concern arises when the same metric fills both roles, as in \citet{guoUltrasoundFrametoVolumeRegistration2023a}, where NCC is used as both loss and evaluation measure. In practice, however, these metrics are rarely reported in isolation, but alongside geometric and, in some studies, parameter-wise measures, providing an intensity-domain perspective that complements the primary accuracy assessment.

\subsection{Robustness and Computational Efficiency}
\label{sec:eval:robustness-efficiency}
The clinical utility of a registration method is determined not only by its accuracy but also by its reliability and speed. The capture range characterises the maximum initial misalignment from which a method can reliably converge to a correct solution; despite its clinical relevance, only eight of the 41 studies report it (Table~\ref{tab:metrics-summary}).
A wide capture range is particularly important in intraoperative settings, where precise initialization cannot be assumed \cite{mohsenisalehiRealTimeDeepPose2019, zhangUnsupervised2D3DDeformable2021,caoNonrigidImagevolumeRegistration2025,yeRobustMethodReal2025,zheng2022unsupervised,gaoFullyDifferentiableFramework2024}. However, capture range definitions and experimental protocols vary substantially across studies, reducing their comparability.

Complementary to the capture range, the success rate, reported by twelve studies, quantifies reliability as the percentage of registration attempts whose final error falls below a clinically acceptable threshold, most often 10\,mm of mTRE \cite{miaoCNNRegressionApproach2016, zheng2022unsupervised, gaoFullyDifferentiableFramework2024} or, more stringently, 1\,mm \cite{gopalakrishnanIntraoperative2D3D2024, gopalakrishnanRapidPatientspecificNeural2025, wangPoseawareDeepPerceptual2026}. 
Inference time is another key factor for real-time or near–real-time applications and the most consistently reported quantity, given by 36 of the 41 studies (Table~\ref{tab:metrics-summary}). Deep learning–based approaches generally achieve substantially faster runtimes than classical optimisation-based methods, often enabling sub-second or millisecond-scale inference \cite{miaoCNNRegressionApproach2016}. Nevertheless, reported runtimes depend strongly on hardware configuration and implementation details, which are not always consistently documented.

\section{Domain Adaptation}
\label{sec:domain-adaptation}
A fundamental challenge in learning-based 2D-to-3D registration is the reality gap between the data used during training and those encountered in clinical practice. Because obtaining large-scale clinical datasets with ground-truth labels is often impractical or even impossible, most works included in this review train on synthetic data \cite{miaoCNNRegressionApproach2016, zhengPairwiseDomainAdaptation2018, gopalakrishnanIntraoperative2D3D2024}. 
However, models trained exclusively on clean synthetic data often fail or underperform when applied to real images in a clinical setting due to clinical noise, varying contrast, and anatomical deformations not captured by the simulation. 

Beyond the synthetic-to-real difficulties, cross-modality differences also introduce a significant domain shift, as the underlying physics acquisition fundamentally differ. These differences result in variations in intensity distributions, texture patterns, and structural appearance. %

To mitigate domain shifts, researchers employ preprocessing strategies and domain adaptation (DA) techniques. DA aims to align feature distributions across domains, whether between synthetic and clinical data or between different imaging modalities, thus improving generalisability.

\subsection{Preprocessing and Simulation Strategies} 
Earlier papers attempted to address this through realistic synthetic data generation, data augmentation, and pre-processing, for example by blending DRRs with real X-ray backgrounds and using Gaussian smoothing to simulate scattering \cite{miaoCNNRegressionApproach2016}.  More recent work has extended this idea with richer augmentation pipelines: \citet{wangContrastiveLearningBased2024} combine multiple noise types, blurring, fogging, and intensity attenuation to simulate low-dose intraoperative X-ray appearance from clean DRRs, as do \citet{zhangEnhancingCNNRegressors2026} with blur and attenuation, while \citet{lecomte-denisDomainAgnostic2D3D2025} adopt the DRR perturbation scheme of \citet{grimm2021pose} and report improved generalization across deformation scenarios. \citet{facenteRealtimeAlignment3D2026} add salt-and-pepper and plasma noise to the DRRs and normalise differing C-arm intrinsics with a homography. 

\subsection{Learnable Domain Adaptation} 
Nine articles implemented dedicated DA modules designed to extract modality-invariant features that look identical to the network regardless of whether the image is real or synthetic. These can be categorized into three main approaches:
\paragraph{Image-to-Image Translation:} These methods use generative models to translate the style of one domain to another. \citet{daiVolumetricTumorTracking2024} utilized a Contrastive Unpaired Translation (CUT-GAN) to add realistic noise and artifacts to synthetic training data. \citet{liDiffusionstimulatedCTUSRegistration} translate in the opposite direction. They simulate physical ultrasound (pUS) images from labelled CT by ray tracing and train a diffusion model to convert real ultrasound images into this pUS style, so that simulated and real inputs share an intermediate representation; a frequency-domain module additionally blends the low-frequency components of simulated and real images. The probe pose is then estimated by a dual-stream self-supervised regression network, trained on CT-derived pUS images and applied at inference to the pUS converted from real ultrasound.
\paragraph{Feature Alignment:} These models force the network to suppress modality-specific style and focus on anatomical content. \citet{zheng2022unsupervised} proposed an Unsupervised Cross-Modality Domain Adaptation Network (UCMDAN) that utilizes adversarial learning to align pixel and feature spaces. \citet{zhengPairwiseDomainAdaptation2018} introduced a Pairwise DA module that uses a multilayer loss to minimize the distance between features extracted from paired real and synthetic images at identical poses. 
\paragraph{Semantic Anchoring:} Some researchers anchor registration in domain-invariant anatomical labels. \citet{kobayashiShapeRegistrationLaparoscopic2025} used semantic organ labels as a shared feature to mitigate the gap, as do \citet{zhangWarmstartedReinforcementLearning2026} with semantic masks and a relative depth map in place of the RGB frame, while \citet{weiDeepLearningApproach2021} focused on vessel structures common to both US and CT/MRI, \citet{wangContrastiveLearningBased2024} and \citet{chiTwostepDeepLearning2023} leveraged contrastive learning and transfer learning, respectively, to adapt general models to specific patient geometries or occluded clinical environments. 

\section{Discussion}
\label{sec:discussion}

\subsection{Architectures}
\label{sec:disc:architectures}

Convolutional neural networks, particularly ResNet variants, remain the dominant backbone across the reviewed literature, appearing in the majority of included studies. Their continued prevalence reflects practical advantages: well-understood training dynamics and modest data requirements compared to Attention-based networks. However, recent work suggests that hybrid architectures combining convolutional feature extraction with attention mechanisms can offer meaningful improvements \cite{quTransformerbased2D3D2024b, wangContrastiveLearningBased2024, khawaledSelfattentionModelRobust2025, xuSVoRTIterativeTransformer2022, daiVolumetricTumorTracking2024}. %
Only \citet{yeRobustMethodReal2025} implemented a fully attention-based solution, as the network handled only 2D inputs, using a combination of Swin-transformers and SE-modules.
These gains do not emerge from replacing convolutions entirely, but from augmenting them with attention. We expect that these hybrid feature extractors will represent the majority of the future 2D-to-3D registration landscape.

In parallel, we found an emerging interest in INRs, which represent the transformation continuously, need no training set and, with periodic activations, allow exact Jacobian-based regularisation \cite{molaei2023implicit} (Section~\ref{sec:arch-inr}). Their per-case optimisation does not yet meet most intraoperative time budgets, but the reviewed work points to a concrete way of closing this gap: starting the fit from the prediction of a population-trained network, which reduced the optimisation from about 13\,min to 197\,s in \citet{qianHybridPopulationbasedPatientspecific2025}. Meta-learned initial weights, which shorten the fitting of coordinate-based networks in other tasks \cite{tancik2021learned}, are a natural next step. INRs also offer a route to the uncertainty gap discussed in Section~\ref{sec:disc:uncertainty}: pairing each INR with a second one that estimates the inverse transformation yields a consensus measure that detected every registration that failed to converge in 4D lung CT \cite{vanharten2024cycle}. We recommend testing both ideas first in offline 2D-to-3D settings, such as limited-angle CBCT estimation, where minutes per case are acceptable.

Foundation models pretrained on large-scale data represent a further opportunity that remains unexplored in the 2D-to-3D setting \cite{vanveldhuizen2025foundationmodelsmedicalimaging}. The broader medical imaging community already uses them as feature extractors \cite{medsam2024, zhang2025biomedclipmultimodalbiomedicalfoundation, ctclip2026}, but none of the reviewed studies does. Given the scarcity of annotated 2D-to-3D data, three integration points follow directly from the pipelines reviewed here. First, in projection-based registration, a single frozen 2D encoder, either general-purpose such as DINOv3 \cite{simeoni2025dinov3} or trained on radiographs such as RAD-DINO \cite{perezgarcia2025raddino}, can embed both the fixed X-ray and the moving DRR, so that the two are compared in the same pretrained feature space; only a light head is trained, an MLP regressing $\varphi_r$ for rigid registration or a decoder for a displacement field. Second, the same frozen features can define the similarity $\mathcal S$ in the test-time optimisation of Eq.~\eqref{eq:tto}, playing the role of the task-trained similarity networks of \citet{gaoFullyDifferentiableFramework2024} and \citet{wangPoseawareDeepPerceptual2026} (Section~\ref{sec:hybrid-tto}) without task-specific training. Third, where the encoder must adapt to a modality or a patient, LoRA \cite{hu2021loralowrankadaptationlarge} adds a trainable low-rank term to the attention projections of Eq.~\eqref{eq:attention}, $W_Q = W_Q^{0} + B_Q A_Q$ and likewise for $W_V$, where $W_Q^{0}$ is the pretrained matrix and $B_Q A_Q$ has rank $r \ll d_k$. The effective projections therefore change, but only the low-rank factors are trained, a small fraction of the parameters, while the pretrained matrices are kept as learned, so the adaptation leaves most of the pretrained representation intact. This suits the patient-specific training used by several reviewed methods. For 3D inputs, volumetric encoders such as CT-CLIP \cite{ctclip2026} can be applied to the preoperative volume once, before the intervention, and so add no intraoperative latency. 

\subsection{Registration Pipeline Design}

The reviewed studies divide into direct and iterative designs, and the choice between them is usually framed as one between the speed of a single forward pass and the accuracy of refinement. As Eq.~\eqref{eq:multipass} shows, the two need not be separate methods. A network trained to predict residual corrections reduces to direct regression for $K=1$ and refines its estimate for $K>1$, so the number of passes can be chosen at inference to fit the time budget of each application. Several reviewed studies already report accuracy as a function of the number of passes. In LXPose, one pass takes 20\,ms, within a 40\,ms control budget, while twelve passes take 187\,ms and reduce the projection error from 0.98 to 0.77\,mm, with accuracy levelling off after about ten passes \cite{facenteRealtimeAlignment3D2026}. SVoRT reduces its mean slice error from 5.99\,mm with one iteration to 4.35\,mm with three \cite{xuSVoRTIterativeTransformer2022}, and the cephalogram registration of \citet{zhangDeformableRegistrationLateral2021} converges after about three passes of roughly 80\,ms each. We therefore recommend designing iterative networks so that every pass returns a valid estimate, and reporting accuracy and run time as functions of $K$, which makes a single method comparable with both direct and iterative baselines.

For applications with more relaxed time constraints, such as preoperative planning or offline treatment verification, the complementarity between DL and classical optimisation is still a compelling approach \cite{gopalakrishnanRapidPatientspecificNeural2025, gopalakrishnan2022fast, downsImprovingAutomaticCerebral2025}. %
As learned cascades and few-iteration methods mature in capacity and robustness, we expect them to progressively absorb the refinement role that classical optimisation currently fills. The hybrid DL-classical paradigm may, therefore, represent an effective but transitional solution rather than a long-term architectural template.

\subsection{Cross-modality Domain Gap}
The gap between synthetic training data and real clinical images remains a central challenge, and its severity varies substantially across problem settings. In X-ray-to-CT registration the forward model is physics-based and well-characterized: DRRs simulate the projective geometry of X-ray formation with reasonable fidelity. The resulting domain gap is primarily of appearance rather than fundamental geometric and structural discrepancy. This partly explains why X-ray to CT methods dominate the literature included in our corpus. Even in this favourable setting, the domain gap is clinically meaningful, and efforts have been made to narrow it down \cite{zhengPairwiseDomainAdaptation2018, zheng2022unsupervised, wangContrastiveLearningBased2024}. 

In a cross-modality setting, where no physics-based forward model exists, this gap widens dramatically. \citet{liDiffusionstimulatedCTUSRegistration} addressed this gap in the CT-US case by combining physics-based US simulation and diffusion models.
Yet, despite these advances (see Section \ref{sec:domain-adaptation}), domain adaptation remains largely underexplored relative to its practical importance.
As the field expands towards more challenging cross-modality pairs (e.g. MRI to US, CT to histopathology), where this gap is vastly wider, robust domain adaptation will become indispensable. One concrete route is already visible in the corpus. Several studies register intermediate representations that look alike in both modalities instead of raw intensities, such as the semantic label maps of \citet{kobayashiShapeRegistrationLaparoscopic2025}, the label images of \citet{oya2D3DDeformable2024} and the vein segmentations of \citet{downsImprovingAutomaticCerebral2025}, so that a network trained on rendered labels can be applied to labels segmented from real images. Promptable segmentation foundation models such as MedSAM \cite{medsam2024} could supply such labels for pairs without a physics-based forward model, for example by segmenting the same structures in the preoperative MRI and in the intraoperative US.

\subsection{Slice-to-Volume Registration}
A notable imbalance in the reviewed literature is the scarcity of slice-based registration (US or MRI slices to 3D volumes), with only 12 studies included. This disparity does not reflect the clinical need. US-guided surgeries are performed more frequently than fluoroscopy-guided ones. This can largely be attributed to the possibility of easily generate high-quality synthetic data from CT scans, allowing training in data-scarce regimes for projection-based registration \cite{qianHybridPopulationbasedPatientspecific2025, gopalakrishnanRapidPatientspecificNeural2025}. 

The key difference is geometric. In projection-based registration, each pixel of an X-ray or DRR integrates attenuation along a ray through the whole volume, so a single projection reflects the entire anatomy, even though the position of a structure along the ray is ambiguous (Section~\ref{sec:eval:geometrical}). A US or MRI slice, by contrast, samples only a thin plane, so the image itself carries no information about the anatomy outside it, and the through-plane position must be recovered from anatomical priors.
This shifts the burden of the registration problem: where projection-based methods can exploit the forward model as a direct source of geometric constraint, slice-based methods depend on learned or atlas-based anatomical knowledge to compensate for the absent through-plane signal.

This dependence shapes the design space of slice-based methods in the corpus. Rather than operating on isolated frames, most studies aggregate information across multiple slices or rely on strong anatomical priors: stack-level processing for fetal brain MRI \cite{xuSVoRTIterativeTransformer2022, khawaledSelfattentionModelRobust2025}, temporal context from US sweeps \cite{guoUltrasoundFrametoVolumeRegistration2023a}, or population-based motion models \cite{mezheritskyPopulationbased3DRespiratory2022}.

This explains why slice-to-volume studies in our corpus tend to be highly application-specific, with limited methodological cross-pollination. The two main strategies have been developed separately, although they address the same problem. SVoRT accumulates context across a stack of MRI slices with a transformer whose positional embedding tells it where each slice lies within its stack \cite{xuSVoRTIterativeTransformer2022}. A freehand US sweep has no such fixed geometry, so the US methods instead estimate the motion between consecutive frames and accumulate context along the sweep \cite{guoUltrasoundFrametoVolumeRegistration2023a, dupuy2D3DDeep2023a}. Feeding these motion estimates to a SVoRT-like transformer in place of the stack and slice indices would let the same attention mechanism accumulate context along a sweep, and would show whether the approach generalises beyond fetal MRI.

\subsection{Evaluation Challenges}
\label{sec:disc:eval}

Cross-study comparison in 2D-to-3D registration is hindered by substantial inconsistencies in evaluation methodologies. Three specific issues require attention.

First, the absence of standardized evaluation protocols means that each new method is typically compared against a different subset of baselines, under different data conditions, making it difficult to reliably compare approaches. Table~\ref{tab:metrics-summary} makes this visible: no single metric is shared by all studies, the most common one, mTRE, appears in 28 of the 41 studies but under two different definitions, the 3D error of Eq.~\eqref{eq:mtre} and the projected error of Eq.~\eqref{eq:mtre-proj}, and only eight studies report a capture range. Community efforts toward shared benchmarks, including standardized datasets, evaluation protocols, and open-source baselines, would substantially accelerate progress.

Secondly, evaluation relies heavily on synthetic data. As Table~\ref{tab:data-extraction-summary} shows, 30 of the 41 studies train exclusively on synthetic data, and 14 evaluate only on synthetic or phantom data; 22 report quantitative results on real clinical images and a further 5 assess real images only qualitatively. Synthetic evaluation enables controlled benchmarking, but it sidesteps the domain gap that constitutes the primary challenge for clinical deployment.

Lastly, we noticed a gap in XAI methods inclusion to validate proper model alignment and interpretability \cite{singh2020explainable}. Out of the 41 included papers, only \citet{khawaledSelfattentionModelRobust2025} and \citet{wangPoseawareDeepPerceptual2026} included an analysis of attention or feature-activation maps, although such maps are debated as a true explanatory tool \cite{bibal2022attention, lopardo2024attention, chung2024evaluating}. This near-complete absence of interpretability is quite striking given that they are intended for a clinical setting. Integrating post-hoc or intrinsic explainability mechanisms will be essential for several aspects such as clinical acceptance, systematic debugging and failure mode identification.

\subsection{Limited Reproducibility}
\label{sec:disc:reproducibility}
As shown in Table \ref{tab:quality-assessment-summary}, the vast majority of the included papers have a limited reproducibility. Few of these studies release both code and trained models, effectively hindering the possibility of ensuring reported results, a central aspect of scientific research, and limiting expansion. 

\subsection{Prediction Uncertainty}
\label{sec:disc:uncertainty}

None of the 41 reviewed methods provide per-case uncertainty estimates alongside their registration predictions. Every method outputs a deterministic point estimate with no indication of confidence. This mirrors the historical situation in classical optimization-based registration, which has been deployed clinically without per-case uncertainty quantification. In both cases, reliability is managed through complementary mechanisms such as visual verification and similarity-based failure detection. This positions uncertainty estimation not as a barrier to clinical use, but it rather represents a missed opportunity.

Treating every prediction as equally trustworthy assumes that registration difficulty is uniform across inputs, which it is not. Registration difficulty varies with anatomies, viewing angles, image quality, and pathology. A featureless soft tissue provides far less information than a bony landmark, and a single-view registration inherently suffers from depth ambiguity. Per-case uncertainty would allow these failure-prone scenarios to be flagged automatically, rather than relying solely on downstream verification.

The broader registration community has begun addressing this gap. \citet{chenSurveyDeepLearning2025} provide a comprehensive taxonomy distinguishing aleatoric uncertainty (arising from data noise, anatomical ambiguity, or lack of distinctive features) from epistemic uncertainty (arising from model limitations and training data scarcity), with practical estimation approaches including Monte Carlo dropout, heteroscedastic variance prediction, and snapshot ensembles. \citet{unberathImpactMachineLearning2021} specifically identified the absence of uncertainty-aware 2D-to-3D registration as a critical gap, noting that the few existing verification approaches provide only binary pass/fail classification rather than spatially resolved confidence maps. 

For real-time clinical applications, heteroscedastic variance prediction offers the most practical path forward: the network predicts both the transformation parameters and an associated variance through a dual-head architecture, adding negligible computational overhead at inference. This approach requires no multiple forward passes making it compatible with the tight latency budgets of intraoperative guidance. Incorporating uncertainty estimation would not transform whether 2D-to-3D registration can be used clinically, but it would expand how it can be used: supporting higher-autonomy pipelines, 
more informative failure handling, and better-calibrated trust between model and clinician.

\subsection{Temporal and Sequential Registration}
\label{sec:disc:temporal}

The clinical reality of most 2D-to-3D registration is fundamentally sequential: fluoroscopy produces continuous video streams, US involves probe sweeps, and fractionated radiotherapy requires repeated inter-session alignment. Yet the overwhelming majority of reviewed methods (37 out of 41) treat each registration as an independent, memoryless inference, discarding all temporal context between frames.

Only four studies explicitly model temporal or spatial continuity \cite{guoUltrasoundFrametoVolumeRegistration2023a, dupuy2D3DDeep2023a, soormallyDeepSensorlessTracking2025, xuSVoRTIterativeTransformer2022}. \citet{guoUltrasoundFrametoVolumeRegistration2023a} combine a frame-to-frame motion estimation network with a frame-to-slice correction module and a similarity-based quality check, demonstrating that without temporal modelling, registration error accumulates rapidly along US video sequences degrading to clinically unacceptable levels within tens of frames. \citet{dupuy2D3DDeep2023a} incorporate the previous registration result and relative probe tracking data as network inputs for prostate biopsy guidance, showing that this temporal prior significantly improves both accuracy and reliability along biopsy trajectories. \citet{soormallyDeepSensorlessTracking2025} track probe orientation along trajectories using spatial context volumes and explicitly study estimation drift over both short and extended sequences.

Temporal modelling brings multiple advantages. The previous time-step registration acts as a strong prior that narrows the search space.  Temporal smoothness acts as a free regularizer, penalizing physically implausible discontinuities. Incremental updates from a nearby initialisation are computationally cheaper than a full re-registration from scratch, benefitting inference time.

The absence of temporal modelling also connects directly to the uncertainty gap identified above: confidence should propagate across frames, with uncertain registrations increasing the uncertainty budget for subsequent predictions rather than being treated as equally reliable starting points. A unified framework that jointly addresses sequential registration and per-frame uncertainty estimation would represent a meaningful advance for more informative clinical systems.

Three families of sequence models can carry this temporal context, with different costs under intraoperative latency budgets, which \citet{facenteRealtimeAlignment3D2026} place at about 40\,ms for closed-loop control and 200\,ms for visual guidance. Recurrent networks such as the LSTM \cite{hochreiter1997lstm} compress the history into a fixed-size state, so each new frame costs the same regardless of sequence length, but their nonlinear recurrence must be trained sequentially and struggles to retain long-range information. Transformers \cite{vaswani2023attentionneed} attend to every frame in a window and train in parallel, but, since every token attends to all others in Eq.~\eqref{eq:attention}, their per-frame cost and memory grow with the length of the attended window. State-space models combine the advantages of both. They carry a hidden state $h_t$ that is updated from the features $e_t$ of frame $t$ by a linear recurrence,
\begin{equation}
\label{eq:ssm}
    h_t = \bar A\, h_{t-1} + \bar B\, e_t,
\end{equation}
where the matrices $\bar A$ and $\bar B$ are obtained by discretising a continuous-time linear system. Because the recurrence is linear, it can be trained in parallel across the whole sequence and run at constant cost per frame at inference \cite{gu2022s4}, and Mamba makes $\bar A$, $\bar B$ and the discretisation step depend on the current input, so that the model learns what to retain \cite{gu2024mamba}. Selective state-space models already process medical and surgical video in this way: EndoMamba scans each endoscopic frame spatially and carries a causal Mamba state from frame to frame for real-time inference on online streams \cite{tian2025endomamba}, Vivim models causal temporal cues in ultrasound and colonoscopy video \cite{yang2024vivim}, and MCITrack uses the Mamba hidden state as a memory for object tracking \cite{kang2024mcitrack}. In registration, by contrast, Mamba has so far been used only along the spatial axis, as a linear-cost alternative to transformer backbones within a single image pair \cite{guo2024mambamorph, wang2024vmambamorph}. A concrete design follows from the pipelines reviewed here: a per-frame registration network, direct or a single refinement pass of Eq.~\eqref{eq:multipass} started from the previous estimate, $\varphi_0^{(t)} = \hat\varphi^{(t-1)}$, whose frame features update a selective state-space memory that is fed back as an additional input. Such a model keeps the per-frame cost of current single-frame methods while adding a learned memory, and its explicit state is also a natural place to propagate the per-frame uncertainty discussed above. Recurrent and windowed-transformer versions of the same design are the natural baselines.

\subsection{Limitations of this Review}
\label{sec:disc:limitations}
This review has several limitations. The search was restricted to PubMed, which covers the clinical and medical-imaging journals in which most 2D-to-3D registration methods are validated but conference proceedings and preprints only in part. Some proceedings, such as those of EMBC, are indexed directly, whereas papers from venues such as MICCAI, CVPR, WACV or ICASSP are retrieved only when an author manuscript has been deposited in PubMed Central, as for three included studies \cite{xuSVoRTIterativeTransformer2022, gopalakrishnanIntraoperative2D3D2024, tianLiftRegLimitedAngle2022}, sometimes long after publication. Even within PubMed, a keyword-based query can miss studies whose titles and abstracts do not use the expected terms, which is why the full query is reported in Appendix~\ref{app:query}. The search covers articles published up to June 2026, in a field where most included studies date from 2024 onwards (Fig.~\ref{fig:year}), and the resulting corpus of 41 studies is too small for finer stratification, for example by anatomy or modality pair, so the trends we report describe this corpus. Finally, we did not pool the reported accuracies, because the studies evaluate on different datasets, largely synthetic (Table~\ref{tab:data-extraction-summary}), and with metric definitions that vary between them (Sections~\ref{sec:eval} and~\ref{sec:disc:eval}); the lack of shared benchmarks that would allow such a comparison is a limitation of the field itself.

\section{Conclusion}
\label{sec:conclusion}
This review of 41 studies demonstrates that deep learning has substantially reshaped 2D-to-3D registration across its diverse clinical applications, from offline volumetric reconstruction to real-time interventional guidance. Across all settings, DL methods offer improved robustness to initialisation, larger capture ranges, and the ability to shift computational burden from inference to training. These advantages are most evident within intraoperative workflows, enabling real-time and robust alignment and achieving millisecond-scale inference that was previously not routinely achievable with traditional optimization methods.

Direct regression remains the most common pipeline, adopted by 23 of the 41 studies for its speed, whereas the remaining 18 refine their estimate over successive steps through repeated network passes, test-time optimisation or per-case fitting, which improves accuracy and capture range at a higher computational cost. Convolutional networks continue to predominate, and the seven transformer-based architectures have all appeared since 2022. Differentiable geometric operators have enabled training without ground-truth transformations, which seven studies adopt exclusively and six combine with supervised objectives. The supporting evidence, however, remains limited in scope. More than half of the studies address X-ray to CT or CBCT registration, and 30 are trained exclusively on synthetic data, so that most reported accuracies reflect performance in simulation; 14 studies do not evaluate on clinical images. Cross-study comparison is further hindered by inconsistent metric definitions, mTRE alone being reported under two different definitions by 28 studies, and by limited reproducibility, the lowest-scoring criterion in our quality assessment (Appendix~\ref{app:quality-assessment}).

Two methodological gaps stand out as most consequential. None of the 41 reviewed methods provide per-case uncertainty estimates, and all but four discard temporal/spatial continuity between frames, ignoring the sequential nature of the targeted workflows. Notably, neither gap prevents clinical use, but closing them would meaningfully expand what 2D-to-3D registration can support: finer-grained trust calibration, failure-aware integration into higher-autonomy pipelines, and exploitation of the strong priors available in sequential imaging. Progress on these two fronts is likely to do more to advance clinical readiness than incremental gains in mean accuracy on synthetic benchmarks.

Reporting a per-case uncertainty estimate alongside each registration would allow guidance systems to identify unreliable results rather than accept them without qualification (Section~\ref{sec:disc:uncertainty}). In sequential imaging, initialising each registration from the preceding estimate and maintaining a learned memory across frames would exploit the strongest prior available, and state-space models offer a constant per-frame cost compatible with intraoperative latency requirements (Section~\ref{sec:disc:temporal}). Given the scarcity of annotated data, pretrained foundation encoders, either frozen or adapted with LoRA, constitute a more data-efficient alternative to encoders trained from scratch (Section~\ref{sec:disc:architectures}). Regarding evaluation, we recommend validation on clinical images and the reporting of 3D errors alongside projected ones, since errors in the image plane may conceal substantial errors in depth. Finally, the release of code and the systematic reporting of capture range would enable comparison on shared benchmarks, which remain unavailable for this task (Sections~\ref{sec:disc:eval} and~\ref{sec:disc:reproducibility}, Appendix~\ref{app:quality-assessment}).

\section*{List of Acronyms}
\phantomsection\label{sec:acronyms}
\begin{description}
  \item[AI] Artificial Intelligence
  \item[AUC] Area Under the Curve
  \item[CBCT] Cone-Beam Computed Tomography
  \item[CMA-ES] Covariance Matrix Adaptation Evolution Strategy
  \item[CNN] Convolutional Neural Network
  \item[CT] Computed Tomography
  \item[CTA] Computed Tomography Angiography
  \item[DA] Domain Adaptation
  \item[DDPM] Denoising Diffusion Probabilistic Model
  \item[DL] Deep Learning
  \item[DoF] Degrees of Freedom
  \item[DRR] Digitally Reconstructed Radiograph
  \item[DSA] Digital Subtraction Angiography
  \item[DSC] Dice Similarity Coefficient
  \item[DVF] Displacement Vector Field
  \item[FDK] Feldkamp-Davis-Kress
  \item[fMRI] functional Magnetic Resonance Imaging
  \item[GAN] Generative Adversarial Network
  \item[GNN] Graph Neural Network
  \item[HD] Hausdorff Distance
  \item[INR] Implicit Neural Representation
  \item[IoU] Intersection over Union
  \item[LoRA] Low-Rank Adaptation
  \item[LRD] Landmark Reprojection Distance
  \item[LSTM] Long Short-Term Memory
  \item[MAE] Mean Absolute Error
  \item[MCD] Mean Contour Deviation
  \item[MeSH] Medical Subject Headings
  \item[MI] Mutual Information
  \item[MI-CLAIM] Minimum Information about Clinical Artificial Intelligence Modeling
  \item[MLP] Multi-Layer Perceptron
  \item[MLS] Moving Least Squares
  \item[MPE] Mean Projection Error
  \item[MRI] Magnetic Resonance Imaging
  \item[MSE] Mean Squared Error
  \item[mTRE] Mean Target Registration Error
  \item[NCC] Normalized Cross-Correlation
  \item[NMI] Normalised Mutual Information
  \item[OAR] Organ at Risk
  \item[PCA] Principal Component Analysis
  \item[PnP] Perspective-n-Point
  \item[PPO] Proximal Policy Optimisation
  \item[PRISMA] Preferred Reporting Items for Systematic Reviews and Meta-Analyses
  \item[ProST] Projective Spatial Transformer
  \item[PSNR] Peak Signal-to-Noise Ratio
  \item[pUS] Physical Ultrasound
  \item[RL] Reinforcement Learning
  \item[RMSE] Root Mean Square Error
  \item[SE] Squeeze-and-Excitation
  \item[SMD] Surface Mesh Deviation
  \item[SSIM] Structural Similarity Index Measure
  \item[SSL] Self-Supervised Learning
  \item[STN] Spatial Transformer Network
  \item[SVoRT] Slice-to-Volume Registration Transformer
  \item[SVR] Slice-to-Volume Registration
  \item[SVT] Slice-Volume Transformer
  \item[Swin] Shifted-Window Transformer
  \item[SyN] Symmetric Normalization
  \item[TRE] Target Registration Error
  \item[UCMDAN] Unsupervised Cross-Modality Domain Adaptation Network
  \item[US] Ultrasound
  \item[XAI] Explainable Artificial Intelligence
\end{description}

\bibliographystyle{plainnat}
\bibliography{references}

\appendix
\section{Quality Assessment}
\label{app:quality-assessment}

The quality of the included studies was assessed using our proposed variant of the MI-CLAIM (Minimum Information about Clinical Artificial Intelligence Modeling) framework, as proposed by \cite{norgeot2020miclaim}. This validated checklist aims to ensure transparency and reproducibility in clinical AI research across six key domains: model scope, data partitioning, model optimization, performance evaluation, model examination, and reproducibility. Each study was scored on a scale of 0 to 2 across nine standardized criteria, allowing for a quantitative comparison of reporting quality.

The classic MI-CLAIM is organized into six functional steps, with multiple criteria. To enable a quantitative comparison of the 41 included studies, we developed a 9-criteria variant that maps the MI-CLAIM requirements relevant to this review into specific, quantifiable engineering domains.
Many of the original framework’s 34 points are designed for general clinical implementation and have been found to be redundant or non-applicable to the developing algorithms in current 2D-to-3D registration studies. By selecting the 9 most relevant criteria, we established a scoring instrument that specifically targets the bottlenecks of the field without irrelevant or administrative metrics. This scope ensured an objective comparison across all 41 studies
\newcommand{\fullmark}{\tikz\fill[black!80] (0,0) circle (0.35em);}        %
\newcommand{\halfmark}{\tikz{                                             %
  \fill[white] (0,0) circle (0.35em);
  \draw[black!80, thick] (0,0) circle (0.35em);
  \fill[black!80] (0,0) -- (0,0.35em) arc (90:270:0.35em) -- cycle;
}}
\newcommand{\emptymark}{\tikz\draw[black!80, thick] (0,0) circle (0.35em);} %

\begin{table*}[tp]
\centering
\caption{Quality Assessment Scores (MI-CLAIM) across the Reviewed Articles.}
\label{tab:quality-assessment-summary}
\small
\setlength{\tabcolsep}{2.5pt}
\begin{tabularx}{\textwidth}{l *{9}{>{\centering\arraybackslash}X} >{\centering\arraybackslash}p{1.2cm}}
\toprule
\textbf{Article} & 
\textbf{\makecell{Model \\ Scope}} & 
\textbf{\makecell{Data \\ Part.}} & 
\textbf{\makecell{Data \\ Prov.}} & 
\textbf{\makecell{Pre- \\ pro.}} & 
\textbf{\makecell{Sel. \\ Opt.}} & 
\textbf{\makecell{Perf.}} & 
\textbf{\makecell{Base. \\ Comp.}} & 
\textbf{\makecell{Model \\ Exam.}} & 
\textbf{\makecell{Repro- \\ duc.}} & 
\textbf{Overall} \\
\midrule
\citet{caoNonrigidImagevolumeRegistration2025} & \fullmark & \fullmark & \fullmark & \fullmark & \fullmark & \fullmark & \fullmark & \fullmark & \fullmark & 18 \\
\citet{chiTwostepDeepLearning2023} & \fullmark & \fullmark & \fullmark & \fullmark & \halfmark & \fullmark & \fullmark & \fullmark & \emptymark & 15 \\
\citet{daiVolumetricTumorTracking2024} & \fullmark & \fullmark & \fullmark & \fullmark & \halfmark & \fullmark & \fullmark & \fullmark & \emptymark & 15 \\
\citet{downsImprovingAutomaticCerebral2025} & \fullmark & \fullmark & \fullmark & \halfmark & \emptymark & \halfmark & \emptymark & \emptymark & \halfmark & 9 \\
\citet{dupuy2D3DDeep2023a} & \fullmark & \fullmark & \fullmark & \fullmark & \halfmark & \fullmark & \fullmark & \emptymark & \emptymark & 13 \\
\citet{facenteRealtimeAlignment3D2026} & \fullmark & \fullmark & \fullmark & \fullmark & \fullmark & \fullmark & \fullmark & \fullmark & \fullmark & 18 \\
\citet{gaoFullyDifferentiableFramework2024} & \fullmark & \fullmark & \fullmark & \fullmark & \fullmark & \fullmark & \fullmark & \fullmark & \fullmark & 18 \\
\citet{gopalakrishnanIntraoperative2D3D2024} & \fullmark & \fullmark & \fullmark & \fullmark & \fullmark & \fullmark & \fullmark & \fullmark & \fullmark & 18 \\
\citet{gopalakrishnanRapidPatientspecificNeural2025} & \fullmark & \fullmark & \fullmark & \fullmark & \fullmark & \fullmark & \fullmark & \fullmark & \fullmark & 18 \\
\citet{guoUltrasoundFrametoVolumeRegistration2023a} & \fullmark & \fullmark & \fullmark & \fullmark & \emptymark & \halfmark & \fullmark & \emptymark & \emptymark & 11 \\
\citet{hindleyOpenSourceDeep2025} & \fullmark & \fullmark & \fullmark & \fullmark & \fullmark & \fullmark & \fullmark & \fullmark & \fullmark & 18 \\
\citet{jiangSemanticallyConsistentDeformable2D3D2025a} & \fullmark & \fullmark & \fullmark & \fullmark & \halfmark & \fullmark & \fullmark & \halfmark & \emptymark & 14 \\
\citet{khamenehAutomaticCNNbased3D2025} & \fullmark & \fullmark & \halfmark & \fullmark & \emptymark & \fullmark & \fullmark & \halfmark & \emptymark & 12 \\
\citet{khawaledSelfattentionModelRobust2025} & \fullmark & \fullmark & \fullmark & \fullmark & \fullmark & \fullmark & \fullmark & \emptymark & \halfmark & 15 \\
\citet{kobayashiShapeRegistrationLaparoscopic2025} & \fullmark & \fullmark & \fullmark & \fullmark & \fullmark & \fullmark & \fullmark & \halfmark & \emptymark & 15 \\
\citet{lecomte-denisDomainAgnostic2D3D2025} & \fullmark & \fullmark & \fullmark & \fullmark & \fullmark & \fullmark & \fullmark & \fullmark & \emptymark & 16 \\
\citet{liDiffusionstimulatedCTUSRegistration} & \fullmark & \halfmark & \fullmark & \fullmark & \fullmark & \fullmark & \fullmark & \emptymark & \emptymark & 13 \\
\citet{mezheritskyPopulationbased3DRespiratory2022} & \fullmark & \fullmark & \fullmark & \fullmark & \emptymark & \fullmark & \fullmark & \fullmark & \emptymark & 14 \\
\citet{mhiriNeuralPatientspecific3D2D2025b} & \fullmark & \fullmark & \fullmark & \fullmark & \emptymark & \halfmark & \fullmark & \emptymark & \emptymark & 11 \\
\citet{miaoCNNRegressionApproach2016} & \fullmark & \fullmark & \fullmark & \fullmark & \fullmark & \halfmark & \fullmark & \fullmark & \halfmark & 16 \\
\citet{mohsenisalehiRealTimeDeepPose2019} & \fullmark & \fullmark & \fullmark & \halfmark & \emptymark & \halfmark & \fullmark & \emptymark & \emptymark & 10 \\
\citet{oya2D3DDeformable2024} & \fullmark & \fullmark & \fullmark & \fullmark & \fullmark & \fullmark & \fullmark & \fullmark & \emptymark & 16 \\
\citet{qianHybridPopulationbasedPatientspecific2025} & \fullmark & \fullmark & \fullmark & \fullmark & \halfmark & \fullmark & \fullmark & \fullmark & \emptymark & 15 \\
\citet{quTransformerbased2D3D2024b} & \fullmark & \fullmark & \fullmark & \fullmark & \halfmark & \halfmark & \fullmark & \fullmark & \emptymark & 14 \\
\citet{soormallyDeepSensorlessTracking2025} & \fullmark & \fullmark & \fullmark & \fullmark & \emptymark & \fullmark & \fullmark & \emptymark & \halfmark & 13 \\
\citet{tianLiftRegLimitedAngle2022} & \fullmark & \fullmark & \fullmark & \fullmark & \halfmark & \fullmark & \fullmark & \halfmark & \fullmark & 16 \\
\citet{wangContrastiveLearningBased2024} & \fullmark & \fullmark & \fullmark & \fullmark & \halfmark & \fullmark & \fullmark & \fullmark & \emptymark & 15 \\
\citet{wangPoseawareDeepPerceptual2026} & \fullmark & \fullmark & \fullmark & \fullmark & \halfmark & \fullmark & \fullmark & \fullmark & \emptymark & 15 \\
\citet{weiDeepLearningApproach2021} & \fullmark & \fullmark & \fullmark & \fullmark & \fullmark & \fullmark & \fullmark & \fullmark & \emptymark & 16 \\
\citet{wu3DIsotropicHighResolution2025} & \fullmark & \fullmark & \fullmark & \fullmark & \fullmark & \halfmark & \fullmark & \fullmark & \fullmark & 17 \\
\citet{wuDVAPRegDualviewAnatomical2026} & \fullmark & \halfmark & \fullmark & \fullmark & \fullmark & \halfmark & \halfmark & \fullmark & \fullmark & 15 \\
\citet{xuSVoRTIterativeTransformer2022} & \halfmark & \fullmark & \halfmark & \fullmark & \emptymark & \fullmark & \fullmark & \halfmark & \halfmark & 12 \\
\citet{yeNonorthogonalKVImaging2025} & \fullmark & \halfmark & \fullmark & \fullmark & \halfmark & \fullmark & \fullmark & \halfmark & \emptymark & 13 \\
\citet{yeRobustMethodReal2025} & \fullmark & \fullmark & \fullmark & \fullmark & \halfmark & \fullmark & \fullmark & \fullmark & \emptymark & 15 \\
\citet{zhangUnsupervised2D3DDeformable2021} & \fullmark & \fullmark & \fullmark & \fullmark & \halfmark & \fullmark & \fullmark & \fullmark & \emptymark & 15 \\
\citet{zhangDeformableRegistrationLateral2021} & \fullmark & \fullmark & \emptymark & \fullmark & \halfmark & \fullmark & \fullmark & \emptymark & \emptymark & 11 \\
\citet{zhangEnhancingCNNRegressors2026} & \fullmark & \emptymark & \halfmark & \halfmark & \emptymark & \fullmark & \halfmark & \halfmark & \emptymark & 8 \\
\citet{zhangUltrasoundSlicetoVolumeRegistration2026} & \fullmark & \fullmark & \fullmark & \fullmark & \fullmark & \fullmark & \fullmark & \halfmark & \emptymark & 15 \\
\citet{zhangWarmstartedReinforcementLearning2026} & \fullmark & \fullmark & \fullmark & \fullmark & \fullmark & \halfmark & \fullmark & \fullmark & \emptymark & 15 \\
\citet{zhengPairwiseDomainAdaptation2018} & \fullmark & \halfmark & \fullmark & \fullmark & \halfmark & \fullmark & \fullmark & \halfmark & \emptymark & 13 \\
\citet{zheng2022unsupervised} & \fullmark & \fullmark & \fullmark & \fullmark & \halfmark & \fullmark & \fullmark & \fullmark & \emptymark & 15 \\
\midrule
\textbf{Average} & \textbf{1.98} & \textbf{1.85} & \textbf{1.88} & \textbf{1.93} & \textbf{1.20} & \textbf{1.78} & \textbf{1.90} & \textbf{1.34} & \textbf{0.56} & \textbf{14.41} \\
\bottomrule
\end{tabularx}
\vspace{2pt}
\begin{flushleft}
\footnotesize
\textbf{Legend:} \fullmark~Full compliance;\quad \halfmark~Partial compliance;\quad \emptymark~Non-compliant or not mentioned. \\
\vspace{0.3em}
\textbf{Criteria:} \textbf{Model Scope}: Model Scope; \textbf{Data Part.}: Data Partitioning; \textbf{Data Prov.}: Data Provenance; \textbf{Pre-pro.}: Preprocessing; \textbf{Sel. Opt.}: Model Selection + Optimisation; \textbf{Perf.}: Performance; \textbf{Base. Comp.}: Baseline Comparison; \textbf{Model Exam.}: Model Examination; \textbf{Reproduc.}: Reproducibility.
\end{flushleft}
\end{table*}

Table \ref{tab:quality-assessment-summary} presents the quality assessment of the 41 included studies.   
By looking at the average score for each category we can quickly notice where the reviewed literature excels and where it is lacking. The most noticeable limitation is towards reproducibility, with just a handful of papers providing the code and weights of the proposed models. Secondly we find a scarce model selection, most papers did not ablate or justify the selection of crucial hyperparameters. Finally we found a scarce model examination, with low inspection of model failure and possible reasons why. 
\section{Research Query}
\label{app:query}
 The research query was formulated on PubMed as follows; its last line restricts the publication date to the period from 1 January 2015 to 30 June 2026:

\begin{lstlisting}
(
    "two-to-three dimensional"[tiab]
    OR
    (
        ( "2D"[tiab]
          OR "two-dimensional"[tiab]
          OR "slice*"[tiab] )
        AND
        ( "3D"[tiab]
          OR "three-dimensional"[tiab]
          OR "volume*"[tiab] )
    )
)
AND
(
    "image registration"[tiab]
    OR "registration"[tiab]
    OR "registering"[tiab]
)
AND
(
    "Artificial Intelligence"[Mesh]
    OR "Machine Learning"[Mesh]
    OR "Deep Learning"[Mesh]
    OR AI[tiab]
    OR "machine learning"[tiab]
    OR "deep learning"[tiab]
    OR "neural network*"[tiab]
    OR CNN[tiab]
    OR "convolutional neural network*"[tiab]
    OR "transformer*"[tiab]
    OR "U-Net"[tiab]
    OR "unsupervised"[tiab]
    OR "supervis*"[tiab]
    OR "self-supervised"[tiab]
    OR "GAN"[tiab]
    OR "generative adversarial"[tiab]
)
AND
("2015/01/01"[dp] : "2026/06/30"[dp])
\end{lstlisting}

\end{document}